\documentclass[epj,referee]{svjour}
\usepackage{graphics}
\usepackage{epsfig}
\usepackage{amssymb}
\usepackage{amsmath}
\usepackage{bbm}
\begin{document}
\title{Nondispersive two-electron wave packets in driven helium}
\author{Peter Schlagheck\inst{1} \and Andreas Buchleitner\inst{2}
}                     
%
%
\institute{Institut f{\"u}r Theoretische Physik, Universit{\"a}t
  Regensburg, D-93040 Regensburg \and Max-Planck-Institut f{\"u}r Physik
  komplexer Systeme, N{\"o}thnitzer Str. 38, D-01187 Dresden} 
\date{Received: \today / Revised version: date}
%
\abstract{
We provide a detailed quantum treatment of the spectral characteristics and of
the dynamics of 
nondispersive two-electron wave packets 
along the 
periodically driven, 
collinear frozen planet 
configuration of helium.
These highly correlated, long-lived wave packets arise as a quantum
manifestation of regular islands in a mixed classical phase space,
which are induced by nonlinear resonances between the external driving 
and the unperturbed dynamics of the frozen-planet configuration.
Particular emphasis is given to the dependence of the ionization rates of the
wave packet states on the driving field parameters and on the quantum
mechanical phase space resolution, preceded by a comparison of 1D and 3D life
times of the unperturbed frozen planet. Furthermore, we study the effect of a
superimposed static electric field component, which, on the grounds of
classical considerations, is expected to stabilize the real 3D dynamics
against large (and possibly ionizing) deviations from collinearity.
} 
\maketitle
\section{Introduction}

As documented by an increasing amount of literature, electronic wave 
packets represent an important object of experimental studies in 
atomic and molecular physics (e.g., \cite{WeiAhnBuc99Nat,CheYea98PRL}).
The possibility to localize the electronic population in phase space 
can be regarded as the crucial link between a quantum wavefunction and 
a classical point particle \cite{Sch26Nw}, and has led to various 
applications in the context of quantum control 
(e.g., \cite{KohO95PRL,AveO96PRL}).
For a long time, it was common belief that the maximum evolution time
until which such a localized phase space distribution can be maintained
is ultimately limited by the time scale corresponding to the
anharmonicity of the underlying classical potential.
After this time scale, any coherent superposition which was initially 
well localized in position and momentum space would have spread out
over the whole classical orbit due to dispersion, i.e., due to the
fact that different components of the superposition contribute with
different classical frequencies.
This is usually termed as the {\em collapse} of the wave packet.

A decade ago, however, it was found that this spreading can actually
be suppressed by means which are common in quantum control, namely by
applying electromagnetic fields to the quantum configuration.
This was explicitly demonstrated for one-electron atoms driven by 
microwave fields with linear \cite{Buc93,DelBuc94AAMOP,BucDel95PRL},
circular \cite{BiaKalEbe94PRL,ZakDelBuc95PRL}, and elliptic polarization
\cite{SacZak99PRA}.
There, a Rydberg wave packet, initially launched along a Kepler orbit
with principal quantum number $n$, is exposed to a
microwave field which is resonant with the
periodic classical motion.
Then, for appropriate field amplitudes (which must neither be too strong
nor too weak, compared to the Coulomb force on the electron) and
for the appropriate relative phase between the microwave field and
the Kepler oscillation, the spreading of the wave packet (which usually
occurs after approx. $n / [3\pi(\Delta n)^2]$ Kepler cycles, given a coherent
superposition of $\Delta n$ unperturbed eigenstates centered around $n$) 
is {\em completely inhibited}.
While perfectly keeping their shape, such {\em nondispersive wave
packets} may well follow the classical motion over time scales up to
$10^6$ Kepler cycles \cite{Buc93,BucDel95PRL,ZakDelBuc95PRL,ZakDelBuc98PRE}
before appreciable ionization (induced by the driving field) sets in
\cite{Buc93,BucDel95PRL,SauBelKoc92PRL}.
This stabilization phenomenon is best understood from the underlying
classical dynamics.
In classical phase space, the time-periodic perturbation of the
microwave destroys the global integrability of the Coulomb dynamics;
local ``islands'' of regular motion, embedded into a ``sea'' of
chaotic, ionizing dynamics, are created by the nonlinear resonance
between the driving field and the unperturbed Kepler motion.
The nondispersive quantum wave packets are nothing but the time-periodic
eigenstates (Floquet states) of the driven atom, which arise from a
local quantization within these resonance islands
\cite{BerZas77PL,HenHol92PRA} (see \cite{BucDelZak02PR} for a recent 
review on the topic).

The above interpretation in terms of nonlinear resonances in a mixed
phase space suggests that the concept of nondispersive wave packets
can be generalized to atomic (or molecular) systems that are more
complex than hydrogen or hydrogen-like Rydberg atoms -- such as, e.g.,
the correlated dynamics in doubly excited helium.
However, such a generalization is not at all obvious, since most
doubly excited two-electron orbits of the bare three-body Coulomb
problem (as the classical version of the doubly excited helium atom)
are unstable and promote rapid autoionization of the configuration.
This prevents the confinement of an initially localized wave packet on
longer time scales.

Nonetheless, extensive studies on the classical dynamics of helium
\cite{RicTanWin93PRA} have identified local regions
of regular motion which are associated with configurations of
particular symmetry of the two-electron atom.
The most prominent and surprising example is the ``frozen planet''
orbit \cite{RicWin90PRL,RicO92JPB}, a highly polarized, nearly collinear
configuration with asymmetric excitations of the electrons. The frozen
planet's stability 
essentially relies on the non-negligible electron-electron 
interaction (the electrons are, contrary to intuition, located on the 
{\em same} side of the nucleus).
Stable eigenmodes sustain radial, vibration-like oscillations of
the outer electron, as well as coupled transverse excursions
of the outer and inner electron. Well-defined eigenfrequencies are
associated with these eigenmodes, what makes the configuration a
good candidate for inducing nonspreading wave packets in the
correlated two-electron dynamics, by near-resonant external driving.

In fact, our earlier analysis of the classical dynamics of helium
in a linearly polarized electromagnetic field already provided strong
support for nondispersive wave packets along the frozen-planet orbit
\cite{SchBuc98JPB,SchBuc99PHD}.
The external perturbation induces regular islands within the phase
space of collinear motion (where electrons and nucleus are lined up parallel
to the driving field polarization axis), which are due to nonlinear resonances
between the drive and the oscillation of the outer electron.
The dynamics within these resonances is in general unstable with
respect to deviations from collinearity (in contrast to the analogous
scenario in atomic hydrogen, the resonantly driven one dimensional
Kepler orbit \cite{LeoRic87JPB}).
However, a static electric field applied parallel to the microwave
polarization axis can be used to enforce stability of the driven
configuration in all three spatial dimensions \cite{SchBuc98JPB}.
This suggests that nondispersive two-electron wave packets can be built on
these stabilized resonances, and, on the basis of semiclassical
considerations, at
experimentally accessible quantum numbers of doubly excited helium
\cite{SchBuc99PHD,Sch99}. Yet, this conjecture remains to be corroborated by
an exact quantum treatment.

With the presently available computational facilities, full blown quantum 
{\it ab initio} calculations of highly correlated two-electron wave
packet states under electromagnetic driving are prohibitive.
However, since the stabilizing 
static field confines the wave packet to the near
vicinity of the field polarization axis, essential properties of these
wave packets appear accessible through the restriction 
of the quantum dynamics along this axis.
Indeed, nondispersive two-electron wave packets associated with the
classical resonances are found in quantum calculations on the
collinear model atom, as we have already reported in a previous letter
\cite{SchBuc99EPL}.
Here we shall give detailed account of our quantum treatment of these objects, 
identify them 
as special eigenstates in the Floquet spectrum of the driven atomic 
system (alike their counterparts in driven hydrogen atoms), and 
investigate their characteristic 
properties.

To make the present contribution self-contained, we 
start out in Section \ref{CL} with a review of the 
classical dynamics of the driven frozen-planet configuration, which was 
already described elsewhere \cite{SchBuc98JPB,SchBuc99PHD}. 
Section \ref{QM} is devoted to the formulation of the quantum
eigenvalue problem for the collinear model.
As a first application thereof, we study the unperturbed collinear atom 
in Section
\ref{1D}, what provides a
remarkable example of regular dynamics in an atomic two-electron system.
In Section \ref{WP}, we show how the nondispersive wave packets emerge
in the spectrum of driven helium (Section \ref{NW}) and discuss their
lifetimes (Section \ref{LW}), as well as their properties in presence of an
additional static
electric field (Section \ref{SF}).

\section{Classical dynamics of the driven frozen-planet configuration}

\label{CL}

\subsection{The frozen-planet configuration of helium}

\label{FP}

In atomic units, which are used throughout this paper, the classical
Hamiltonian of the electromagnetically driven helium atom reads
\begin{eqnarray}
H & = & \frac{{\bf p}_1^2}{2} \; + \; \frac{{\bf p}_2^2}{2} 
  \; - \; \frac{Z}{|{\bf r}_1|} \; - \; \frac{Z}{|{\bf r}_2|}
  \; + \; \frac{1}{|{\bf r}_1 - {\bf r}_2|} \nonumber \\
 && \; + \; {\bf F}(t) \, ( {\bf r}_1 + {\bf r}_2 ), \label{hcl}
\end{eqnarray}
with ${\bf r}_i = (x_i, y_i, z_i)$ and 
${\bf p}_i = (p_{ix}, p_{iy}, p_{iz})$ the position and the momentum of
electron $i = 1,2$, respectively, $Z = 2$ the nuclear charge, and
\begin{equation}
{\bf F} \; = \; ( F \cos \omega t + F_{st} ) \, {\bf e}_z \label{F}
\end{equation}
the external, linearly polarized driving at 
frequency $\omega$ and amplitude
$F$, possibly superimposed by an additional, static component with
field strength $F_{st}$ (${\bf e}_z$ represents the unit vector along
the $z$ axis).
In analogy to driven hydrogen \cite{LeoPer78PRL} as well as to the
unperturbed helium atom \cite{Per77ACP}, the Hamiltonian (\ref{hcl})
exhibits general scaling laws: The classical dynamics generated by
(\ref{hcl}) remain invariant if all variables and parameters of the
system are transformed according to
\begin{eqnarray}
{\bf r}_i & \longmapsto & N^2 \, {\bf r}_i \quad (i = 1,2),
\label{scr} \\
{\bf p}_i & \longmapsto & N^{-1} \, {\bf p}_i \quad (i = 1,2), \nonumber \\
t & \longmapsto & N^3 \, t, \nonumber \\
{\bf F} & \longmapsto & N^{-4} \, F, \nonumber \\
\omega & \longmapsto & N^{-3} \, \omega, \nonumber \\
H & \longmapsto & N^{-2} \, H, \nonumber 
\end{eqnarray}
where $N$ represents an arbitrary, real positive quantity.
Since any classical action scales linearly with $N$ (as does the
product of position and momentum), we shall identify $N$ with the
principal quantum number of the inner electron, which corresponds to
the action variable of its Kepler orbit.
Due to the above scale invariance, we can restrict our classical
phase space analysis to a fixed quantum number $N = 1$, and use 
(\ref{scr}) to deduce the actual phase space structure of
the energy range of interest.

\begin{figure*}
\begin{center}
\leavevmode
\epsfxsize15cm
\epsfbox{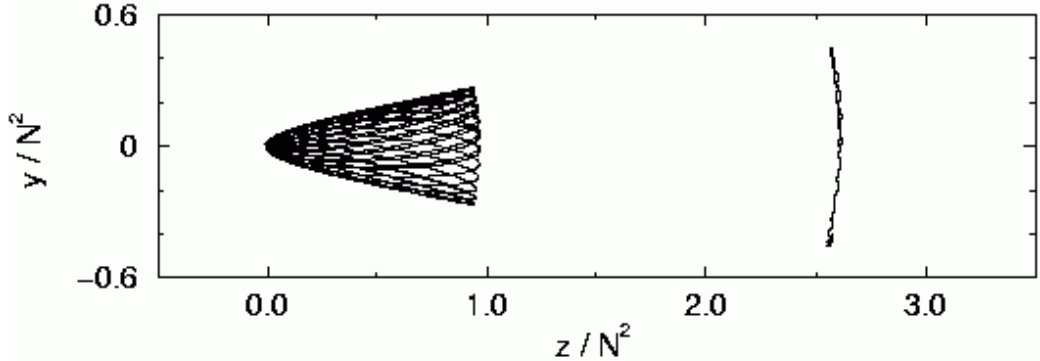}
\end{center}
\caption{A typical (planar) trajectory of the frozen-planet configuration.
 While the inner electron
  oscillates on eccentric Kepler trajectories around the nucleus, the
  outer electron is dynamically stabilized and remains nearly
  ``frozen'' around a given equilibrium distance. The configuration is
  stable against autoionization and represents a large regular region
  in the classical phase space of helium. 
\label{fp2d}}
\end{figure*}

Figure \ref{fp2d} shows a solution of the classical equations of
motion generated by (\ref{hcl}), which illustrates the frozen planet
configuration we shall focus on throughout this paper.
Both electrons are located on the {\em same} side of the nucleus, librating
with respect to a common symmetry axis. 
While the inner electron oscillates on extremely eccentric Kepler
trajectories around the nucleus, the outer electron is dynamically
stabilized due to the rapid oscillation of the inner, and remains
nearly ``frozen'' in the vicinity of an equilibrium distance where the
attractive and repulsive forces due to the other charged particles
cancel each other when averaged over one Kepler cycle of the inner
electron.
The configuration is {\em classically stable} against autoionization
and defines a relatively large region of regular motion in the
dominantly chaotic phase space of helium.
Exact quantum calculations on doubly excited helium \cite{RicO92JPB}
have indeed revealed the existence of long-lived autoionizing states (the
energetically lowest one, 
below the $N = 3$ threshold)  which
are localized along the frozen-planet orbit.

\begin{figure*}
\begin{center}
\leavevmode
\epsfxsize15cm
\epsfbox{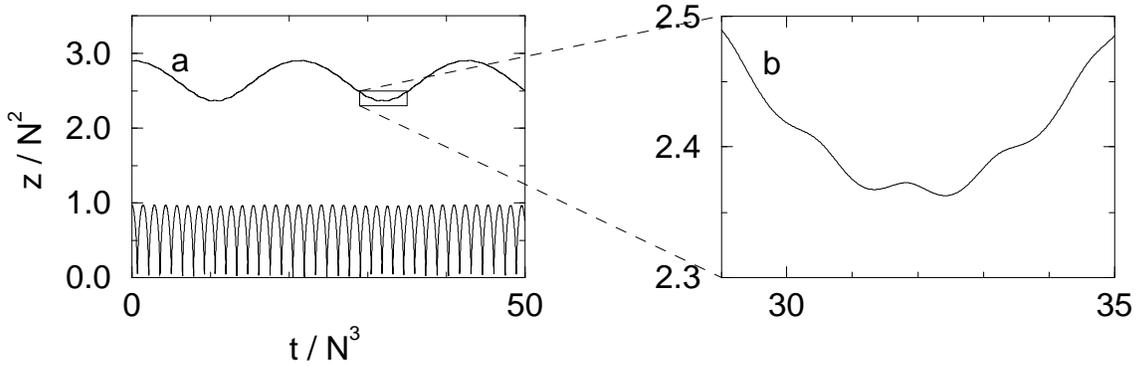}
\end{center}
\caption{Collinear trajectory of the frozen-planet configuration. The
  positions of the electrons are plotted as a function of time. We see
  that the dynamics is characterized by two almost separable modes:
  (i) the Kepler mode of the inner electron, which also manifests in
  the motion of the outer electron, as shown in the magnification
  (b) of the latter's trajectory; (ii) the slow oscillation of the outer
  electron around the equilibrium distance, which induces a feeble
  (and, at this scale, hardly visible) modulation of the amplitude of
  the inner electron's excursion.
\label{fp1d}}
\end{figure*}

For a suitable choice of initial conditions, the motion of the electrons
remains confined to the $z$ axis, what defines the subspace of collinear
motion. In this simplest case,
the regular
frozen-planet dynamics is characterized by two modes which almost
perfectly separate (Fig.~\ref{fp1d}): the fast Kepler mode of the
inner electron (which is also manifest in the motion of the outer
electron, see Fig.~\ref{fp1d}(b)) and the slow oscillation of the
outer electron around the equilibrium distance (which also affects
the motion of the inner electron, as a slow modulation of its maximal
excursion). 

\begin{figure*}
\begin{center}
\leavevmode
\epsfxsize15cm
\epsfbox{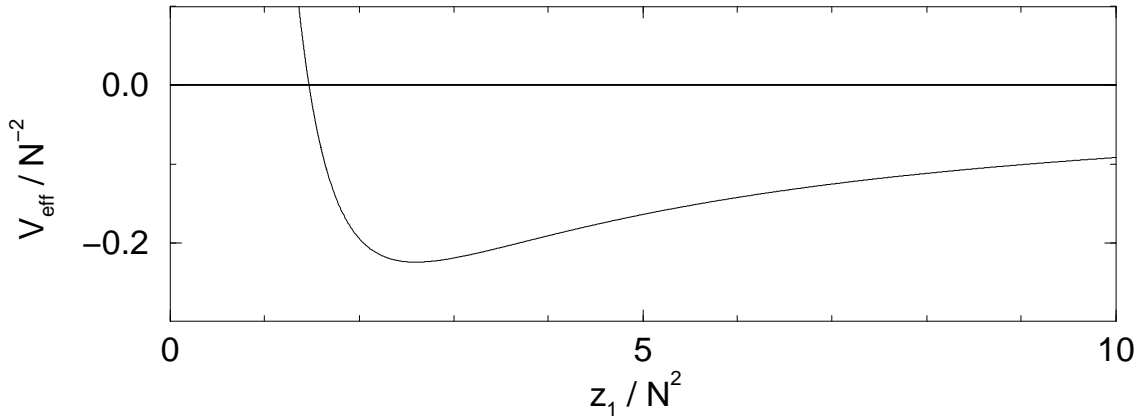}
\end{center}
\caption{Effective adiabatic potential describing the slow oscillation
  of the outer electron around its equilibrium position (see
  \protect\cite{OstPru95JPB}). Intrinsic scales for frequency and
  field strength can be infered from the shape of the potential. 
  The natural scale of the field strength, $F_I \simeq 0.03 \,
  N^{-4}$, is given by the maximum slope of the potential (at $z_1
  \simeq 3.7 \, N^2$). 
  The frequency scale, $\omega_I \simeq 0.3 \, N^{-3}$, is given by the
  curvature of the potential at its minimum at $z_1 \simeq 2.6 \,
  N^2$, i.e., by the frequency of small oscillations around the
  equilibrium.
\label{adiab}}
\end{figure*}

This separation of time scales (the Kepler oscillation is almost 15
times faster than the slow mode of the outer electron) allows to treat
the frozen-planet dynamics within the formalism of adiabatic
invariants \cite{LicLie}.
This defines an effective potential which describes the slow dynamics
of the outer electron, in the combined field of the nucleus and of the
rapidly oscillating inner electron \cite{OstPru95JPB}.
As we see in Fig.~\ref{adiab}, this potential is attractive
Coulombic for large distances (due to the attraction by the nucleus),
strongly repulsive for short distances (due to encounter with the
inner electron) and exhibits a local minimum at the equlibrium
distance.

From the shape of this potential, we can immediately infer intrinsic
scales for the frequency and the field strength, which depend only on
the inner electron's quantum number $N$, and which completely determine
the effect of an external drive on the configuration. 
The natural scale for the field strength, $F_I \simeq 0.03 \, N^{-4}$,
is given by the maximum slope of the potential (at $z_1 \simeq 3.7 \,
N^2$; throughout the paper, subscript 1 denotes the outer, and subscript 2 the
inner electron), and indicates the maximum static field that can be
applied to the configuration without ionizing it.
The frequency scale $\omega_I \simeq 0.3 \, N^{-3}$ is given by the
curvature of the potential at its minimum, i.e., by the frequency of
small oscillations around the equilibrium position.
In the following, we shall study the effect of an electromagnetic field 
which {\em resonantly} drives the outer electron's slow oscillation, 
i.e., a field with amplitude $F < F_I$ and frequency 
$\omega \simeq \omega_I$.

\subsection{Phase space of the collinear, driven configuration}

\label{PC}

In the presence of the time-periodic perturbation, the dynamics of the
collinear, driven frozen-planet evolves 
in a five-di\-men\-sio\-nal phase space spanned by the positions
and momenta of the electrons, and by the phase $\omega t$ of the
driving field.
A complete visualization of the driven dynamics within a simple
two-dimensional Poincar\'{e} surface of section is therefore not
possible.
However, for $\omega \simeq \omega_I$, and for field amplitudes $F <
F_I$ which do not appreciably affect the Kepler motion of the
inner electron, the separation of time scales allows to map the
phase space structure onto a two-dimensional surface by means of a
two-step Poincar\'{e} section method \cite{SchBuc99PHD,Sch99}.
We begin with a simple Poincar\'{e} section by plotting the outer
electron's phase space variables $z_1$, $p_{z1}$ whenever the inner
electron reaches the nucleus ($z_2 = 0$) -- i.e., whenever the phase 
of the Kepler oscillation assumes a fixed value.
In the time domain, the points obtained by this section are separated by
the Kepler period which is much shorter than any other time scale in the
system.
The dynamics generated by the {\em slow} modes -- i.e., the outer 
electron's oscillation within the effective potential, as well as the 
external driving -- is therefore very well monitored by this 
sequence of points.

Neglecting the effect of resonances between fast and slow dynamics,
the intersection of the Lagrangian manifold with the surface defined 
by $z_2 = 0$ can now be very well approximated by a cubic interpolation 
between subsequent points in the above sequence.
This yields a {\em continuous} trajectory which can be used to perform
a second Poincar\'{e} section, by fixing the phase of the driving
field $\omega t = \phi_0$ (mod $2 \pi$).
As a consequence, we effectively plot the outer electron's position
and momentum for $z_2 = 0$ and for fixed $\omega t$.
The reduction of the phase space to a two-dimensional surface is
completed by restricting the initial conditions to those with fixed
value $N = 1$ of the inner electron's action.
This latter quantity represents an adiabatic invariant of the system
and therefore remains nearly constant as time evolves.

\begin{figure*}
\begin{center}
\leavevmode
\epsfxsize15cm
\epsfbox{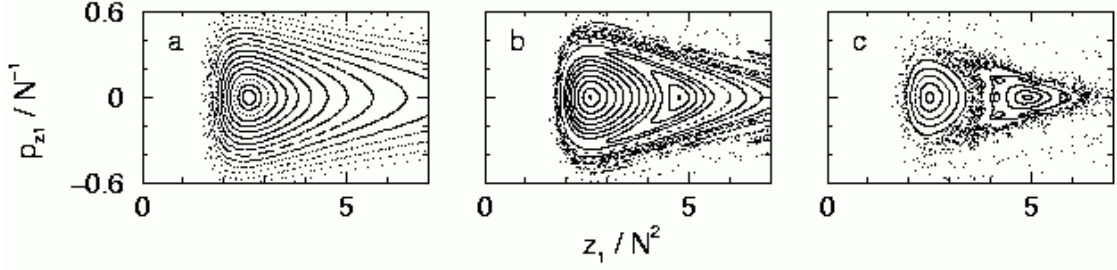}
\end{center}
\caption{Phase space of the collinear, driven frozen-planet
  configuration, for variable field amplitudes, (a) $F = 0$, 
  (b) $0.001 \, N^{-4}$, (c) $0.005 \, N^{-4}$, at fixed
  driving frequency $\omega = 0.2 \, N^{-3}$. 
  The momentum $p_{z1}$ of the outer electron is plotted versus its
  position $z_1$ at fixed action $N = 1$ of the Kepler mode, fixed
  position $z_2 = 0$ of the inner electron, and fixed phase 
  $\omega t = 0$ (mod $2 \pi$) of the driving field
  \protect\cite{SchBuc99PHD,Sch99}. 
  The regular phase space of the unperturbed atom (a) turns mixed
  regular-chaotic in presence of the external perturbation (b, c).
  The nonlinear 1:1 resonance between the driving field and the
  unperturbed oscillation of the electron gives rise to a substructure
  at $F = 0.001 \, N^{-4}$ (b) (centered around $z_1 \simeq 4.8 \,
  N^2$), which transforms into a separate regular island within the
  chaotic sea at larger field amplitude $F = 0.005 \, N^{-4}$ (c). 
\label{pspF}}
\end{figure*}

Fig.~\ref{pspF} shows the Poincar\'{e} section 
that is obtained by the two-step method described above, for
fixed frequency $\omega = 0.2 \, N^{-3} < \omega_I$, fixed driving
phase $\omega t = 0$, and for variable field amplitude $F$.
In the unperturbed case ($F = 0$, Fig.~\ref{pspF}(a)), we recognize
a regular phase space structure, with closed curves corresponding to
the regular oscillation of the outer electron within the effective
potential. 
If we switch on the driving field, phase space turns mixed
regular-chaotic.
The bounded frozen-planet dynamics is now represented by a local
regular region centered around the equilibrium point, outside which
the dynamics is chaotic and leads to ionization.
This remaining regular region is, due to the Kolmogorov-Arnold-Moser
(KAM) theorem \cite{LicLie}, still large for weak field amplitudes
($F = 0.001 \, N^{-4}$, Fig.~\ref{pspF}(b)), but decreases in size with
increasing $F$.
As a consequence of the Poincar{\'e}-Birkhoff theorem \cite{LicLie},
elliptic substructures, induced  by nonlinear resonances between the
external driving and the unperturbed oscillation of the outer electron,
emerge within the regular region.
The prominent substructure located around $z_1 \simeq 4.8 \, N^2$ for
$F = 0.001 \, N^{-4}$ (Fig.~\ref{pspF}(b)) corresponds to the 1:1
resonance where one oscillation cycle of the outer electron is
completed after precisely one period of the driving field.

At larger field amplitude ($F \simeq 0.003 \, N^{-4}$ for the
frequency used in Fig.~\ref{pspF}), the field-induced resonance
eventually disconnects from the rest of the regular domain.
The phase space then exhibits two separate, large regular islands
embedded into the chaotic sea (see Fig.~\ref{pspF}(c) at $F = 0.005 \,
N^{-4}$): the {\em intrinsic} island (centered around $z_1 \simeq 2.5
\, N^2$ in Fig.~\ref{pspF}(c)) which is essentially due to the
intrinsic nonlinear frozen-planet dynamics with the external field
acting only as a small perturbation, and the field-induced {\em 1:1
  resonance} island (centered around $z_1 \simeq 4.9 \, N^2$ in
Fig.~\ref{pspF}(c)) which arises from the combination of two equally
important nonlinear components: the coupling to the external driving
field and the internal interactions between the charged particles of
the atom.

\begin{figure*}
\begin{center}
\leavevmode
\epsfxsize15cm
\epsfbox{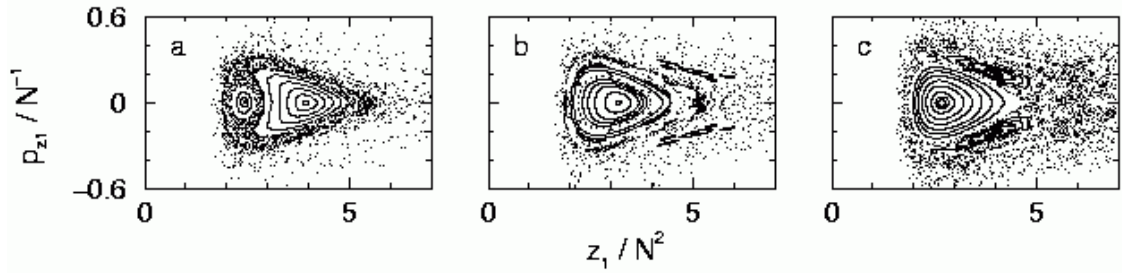}
\end{center}
\caption{Phase space of the collinear, driven frozen planet configuration for
  variable driving frequency, (a) $\omega = 0.25 \, N^{-3}$, (b) $0.3 \,
  N^{-3}$,  (c) $0.4 \, N^{-3}$, at fixed amplitude $F
  = 0.005 \, N^{-4}$ and phase $\omega t = 0$ of the drive.
  With increasing $\omega$, the 1:1 resonance island (centered around
  $z_1 \simeq 3.9 \, N^2$ in (a)) shifts towards the equilibrium
  distance, and finally undergoes a smooth transition into the
  intrinsic island at $\omega = \omega_I = 0.3 \, N^{-3}$ (b).
  The two islands at $z_1 \simeq 4 \, N^2$, $p_{z1} \simeq \pm 0.2 \,
  N^{-1}$ for $\omega = 0.4 \, N^{-3}$ (c) are induced by the 2:1
  resonance where one period of the outer electron's motion
  matches two field cycles.
\label{pspO}}
\end{figure*}

Fig.~\ref{pspO} illustrates the effect of tuning 
the frequency $\omega$ 
at fixed field amplitude $F = 0.005 \, N^{-4}$. 
With increasing $\omega$, the field-induced resonance shifts
towards lower excitations within the effective potential, i.e.,
towards unperturbed orbits with higher frequencies. 
Consequently, the resonance island is located closer (as compared to
Fig.~\ref{pspF}(c)) to the
equilibrium distance for $\omega = 0.25 \, N^{-3}$
(Fig.~\ref{pspO}(a)) and finally undergoes a smooth transition into
the intrinsic island at $\omega = 0.3 \, N^{-3} \simeq \omega_I$
(Fig.~\ref{pspO}(b)). 
Above the intrinsic frequency $\omega_I$ -- which corresponds to the
maximum frequency of free oscillations within the effective
potential -- the electromagnetic field can no longer induce a 1:1
resonance.
Resonances of higher order, however, are still possible (e.g., the 2:1
resonance, where one period of the outer electron is
completed precisely after two field cycles; the elliptic islands created by  
this 
resonance are located at $z_1 \simeq 4 \, N^2$ and $p_{z1} \simeq \pm
0.2 \, N^{-1}$ in Fig.~\ref{pspO}(c)). 

\begin{figure*}
\begin{center}
\epsfxsize15cm
\epsfbox{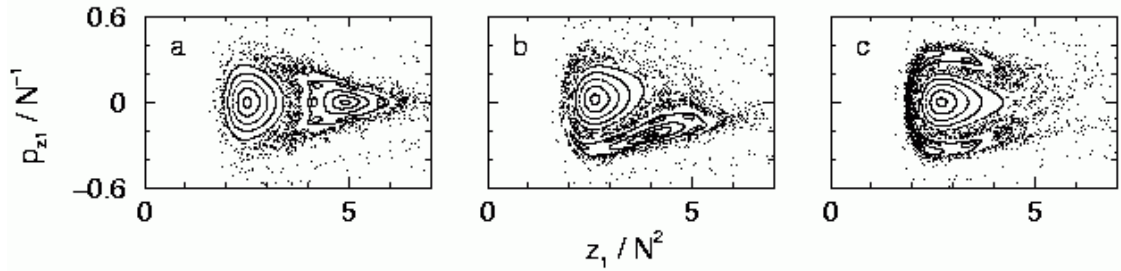}
\end{center}
\caption{Phase space of the collinear, driven configuration for
  variable phases of the driving field, 
(a) $\omega t = 0$, (b) $\pi / 2$, 
  (c) $\pi$, at fixed frequency $\omega = 0.2 \, N^{-3}$
  and field amplitude $F = 0.005$.
  While the intrinsic island remains basically at rest during time
  evolution, the field-induced 1:1 resonance island faithfully traces
  the resonantly driven trajectory of the outer electron.
\label{psp_rot}}
\end{figure*}

The crucial qualitative difference between the intrinsic island and
the field-induced resonances is best illustrated by visualizing the
phase space structure for different times, i.e., by performing the above
two-step Poincar\'{e} section for different phases $\omega t$ of the
driving field.
Fig.~\ref{psp_rot} shows the phase space of the driven configuration
at $\omega = 0.2 \, N^{-3}$, $F = 0.005 \, N^{-4}$, for the field
phases $\omega t = 0$, $\pi / 2$, and $\pi$. 
We see that the intrinsic island remains basically at rest during the
time evolution. The field-induced 1:1 resonance island, on the other
hand, oscillates once around the intrinsic island within one field
cycle.
 
\subsection{Stabilization against deviations from collinearity}

\label{SD}

\begin{figure*}
\begin{center}
\leavevmode
\epsfxsize15cm
\epsfbox{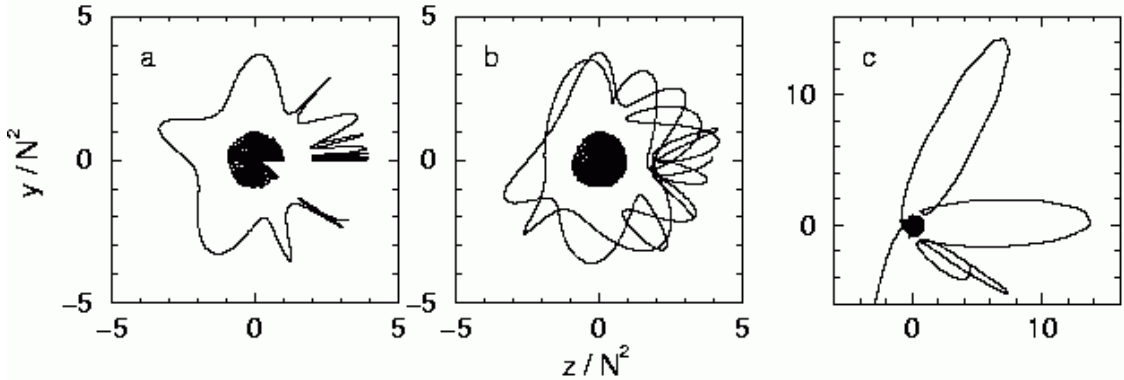}
\end{center}
\caption{Unstable trajectory of the driven frozen-planet configuration 
  in two-dimensional configuration space. 
  The configuration is initially launched at the center of the
  collinear 1:1 resonance island for $\omega = 0.25 \, N^{-3}$ and $F
  = 0.005 \, N^{-4}$ (see Fig.~\protect{\ref{pspO}}(a)), except for a
  small transverse component $y_1 = 0.01 \, N^2$ of the outer
  electron's position.
  The resulting time evolution of the trajectory is depicted from 0 to
  15 field cycles in (a), from 15 to 30 field cycles in (b), and from
  75 to 91 field cycles in (c).
  We see that the initially small deviation of the configuration from 
  the $z$ axis increases with time, until, after approx.~$10$ field cycles, 
  both electrons flip over to the other side of the nucleus (a).
  The configuration then performs several chaotic rotations around 
  the nucleus (b), before the correlation between the electrons eventually 
  breaks down and the atom ionizes (here at about 90 field cycles) (c).
\label{unstab}}
\end{figure*}

In contrast to their counterparts in driven hydrogen
\cite{BucDelZak02PR,BucDel97PRA}, however, the resonance islands of
the collinear, driven frozen-planet dynamics are {\em not} embedded
in a regular phase space domain of the three-dimensional
problem. 
Fig.~\ref{unstab} shows the evolution of the configuration launched at the
center of the 1:1 resonance island of Fig.~\ref{pspO}(a) (i.e., for
$\omega = 0.25 \, N^{-3}$ and $F = 0.005 \, N^{-4}$) if the outer
electron is slightly displaced from collinearity by a small transverse
component $y_1 = 0.01 \, N^2$ of its position.
Due to the effective torque exerted by the driving field on the 
configuration, the initially small deviation from the field polarization 
axis increases rapidly with time, until, after about ten field cycles, 
both electrons ``flip over'' to the other side of the nucleus.
The configuration then performs several chaotic rotations around the 
center, until the correlation between the electrons eventually breaks 
down and the atom ionizes after a subsequent electron-electron collision.
This transverse instability turns out to be characteristic not only for 
the 1:1 resonance, but also for the intrinsic island, as well as for
the 2:1 and for higher resonances (exceptions are tiny
resonances of higher order, such as the 3:2 resonance, see
\cite{SchBuc99PHD,Sch99}).
To launch nondispersive wave packets along the frozen-planet orbit of
the real three-dimensional atom, an additional perturbation of the
atom is therefore required, in order to stabilize the electronic motion
with respect to deviations from collinearity.

\begin{figure*}
\begin{center}
\leavevmode
\epsfxsize15cm
\epsfbox{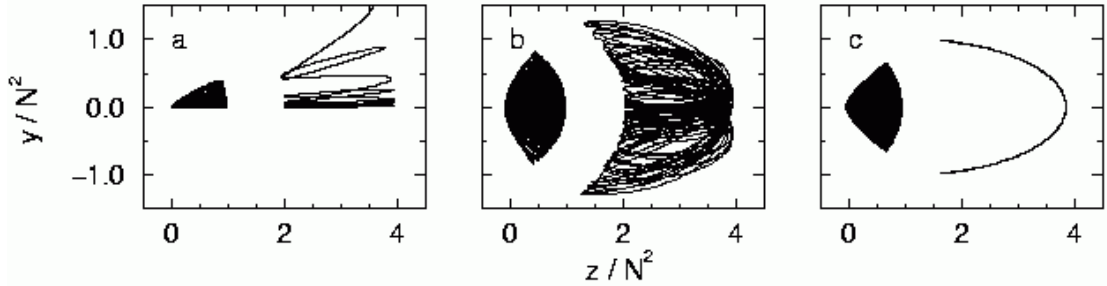}
\end{center}
\caption{Stabilization of the transverse frozen planet dynamics at
  $\omega = 0.25 \, N^{-3}$ and $F = 0.005 \, N^{-4}$.
  As in Fig.~\protect{\ref{unstab}}, the configuration is initially 
  launched at the center of the collinear 1:1 resonance island, except
  for a small transverse component $y_1 = 0.01 \, N^2$ of the outer
  electron's position.
  Without the static field (a), the electrons drift away from the $z$
  axis and enter into chaotic rotations around the nucleus.
  A static electric field, $F_{\rm st}=0.1F$, along the $z$ axis, directed such
  as to force the electrons into the positive $z$ direction,
  counterbalances this drift mechanism (b); the electrons are
  driven back towards the $z$ axis, and the configuration performs
  stable, quasi-periodic oscillations around the fundamental orbit
  shown in (c).
\label{stab}}
\end{figure*}

In fact, such a stabilization can be achieved by adding a {\em static} 
component $F_{st}$ to the electromagnetic field (see Eq.~(\ref{F})).
In resonantly driven hydrogen, a static field, applied parallel 
to the microwave polarization axis and forcing the electron away from
the nucleus, tends to confine the electron to the immediate vicinity
of the polarization axis
\cite{BucDelZak02PR,LeoRic87JPB,SacZakDel98EPD}.
This confinement effect can now be imported to stabilize the motion 
of the driven frozen-planet -- by preventing the configuration from
chaotic rotations.
Indeed, for weak static fields, $F_{st} \simeq 0.1 F$, and for not too
strong driving amplitudes $F$ (e.g., $F = 0.005 \, N^{-4}$ at $\omega
= 0.25 \, N^{-3}$, see Fig.~\ref{stab}), small deviations from
collinearity do {\em not} lead to ionization.
Instead of drifting away to the other side of the nucleus (as in the
absence of the static field, Fig.~\ref{stab}(a)), the electrons are
forced back towards the field polarization axis (Fig.~\ref{stab}(b)), 
and the configuration performs stable, quasi-periodic oscillations 
around the fundamental orbit shown in Fig.~\ref{stab}(c).

This stabilization 
mechanism permits to embed the nonlinear resonances into regular
regions within the unrestricted phase space of three-dimensional
motion, while the electronic dynamics within the collinear subspace is only
mar\-gi\-nal\-ly modified by the weak static field.
We find that this multidimensional phase space region is particularly
large for the 2:1 resonance (Fig.~\ref{pspO}(c)):
By means of the semiclassical Einstein-Brillouin-Keller (EBK)
quantization criterion \cite{Kel58AP} (which typically yields rather
pessimistic results, see \cite{MueBurNoi92PRA}), we
estimate that the minimum principal quantum number of the inner
electron needed to fully localize a quantum state on the stabilized
2:1 resonance island lies in the range $N \simeq 50 \ldots 100$
\cite{SchBuc99PHD,Sch99}.
Such double excitations are clearly beyond reach in state-of-the-art
experiments on helium, which currently can excite $N \simeq 10$ \cite{DomO96PRA}.
However, they should be accessible in earth alkaline atoms such as
barium, where multi-step isolated-core excitations \cite{CooGal78PRL}
permit selective transitions up to states with quantum numbers $N \simeq
100$ of the inner electron \cite{EicLanSan90PRL}.
At $N = 50$, the field parameters that are required to generate the 
2:1 resonance islands of Fig.~\ref{pspO}(c) are given by 
$\omega / ( 2 \pi ) \simeq 20$ GHz and $F \simeq 8$ V$/$cm.
Hence, by virtue of this semiclassical argument, 
nondispersive two-electron wave
packets appear realizable in earth alkalines, at 
microwave frequencies and field
amplitudes commonly used in state-of-the-art ionization
experiments on one-electron atoms \cite{Koc95PsD}.

\section{Formulation of the quantum eigenvalue problem}

\label{QM}

\subsection{The collinear Hamiltonian}

The preceding classical analysis strongly suggests the existence of
nondispersive wave packets which propagate along the frozen-planet
orbit, together with a rough estimate of the minimum required 
quantum number for
their unambiguous identification in the spectrum of doubly excited helium.
Reliable information on the quantum properties of the field-induced
resonances, however, can only be obtained by quantum calculations.
In particular, classical mechanics tells nothing about the lifetimes
of the wave packets, since these are associated with a classically
forbidden tunneling process. 
It is, on the other hand, also clear that a full-blown quantum
treatment of such highly correlated states under external driving is
still beyond reach of the presently available computational facilities.
For {\em unperturbed} helium, where the
total angular momentum is conserved, autoionizing states can
presently be calculated for quantum numbers of the inner electron
reaching $N \simeq 15 \ldots 20$
\cite{RicO92JPB,GreDel97EPL}.
Therefore, in the near future 
there is not much hope for an accurate treatment of the
periodically {\em driven} problem, with a large number of angular momenta
coupled by the field, at comparable or even higher
excitations.

Nevertheless, essential {\em qualitative} information on the wave 
packets can be obtained by an approximate quantum description, which is 
motivated by the classical motion of the electrons.
As discussed in the previous section, the static field confines the
electrons to the vicinity of the field polarization axis 
(Fig.~\ref{stab}), without major modifications of the dynamics in the
collinear subspace.
Hence, also a quantum state associated with the transversly stabilized
wave packet 
orbit should exhibit strong localization close to the
$z$ axis.
We therefore expect that essential properties of such a quantum state
are reproduced by a {\em collinear} model which restricts the
electrons to move along the field polarization axis.
This, however, requires that the model takes into account the {\em full}
Coulomb interaction between the charged particles (i.e., that it really 
represents the exact quantum analog of the classical, collinear
configuration).
A smoothing of the Coulomb singularity, which is frequently employed in
one-dimensional models of driven atoms (e.g., \cite{GroEbe92PRL}), may
not be permitted, since it allows the electrons to penetrate to the
other side of the nucleus, what inevitably destroys the frozen-planet
configuration.

Consequently, we write the Hamiltonian that generates the quantum dynamics of
the driven collinear configuration as
\begin{eqnarray}
H & = & - \frac{1}{2} \frac{\partial^2}{\partial z_1^2} 
  \; - \; \frac{1}{2} \frac{\partial^2}{\partial z_2^2} 
  \; - \; \frac{Z}{z_1} \; - \; \frac{Z}{z_2} \; + \; \frac{1}{z_1 - z_2} 
  \nonumber \\
&& - \frac{F}{\omega} \, \sin \omega t \left( \frac{1}{\rm i} 
  \frac{\partial}{\partial z_1} + \frac{1}{\rm i} 
  \frac{\partial}{\partial z_1} \right) \; - \; F_{st} \, ( z_1 + z_2
  ), \label{Hqm}
\end{eqnarray}
where the {\em exact} Coulomb interactions between the charged particles 
are taken into account.
Here, $z_1$ and $z_2$ represent the {\em Cartesian} (not spherical)
coordinates of the electrons along the field polarization axis, with
$0 < z_2 < z_1 < \infty$ -- i.e., both electrons are located on the
same side of the nucleus, with electron 2 closer to it than electron
1.
Note that the electrons effectively appear as {\em distinguishable}
particles.
This ansatz is consistent with the restriction to collinear motion,
since in the classical collinear configuration the electrons are not able 
to pass each other and to exchange their positions.
Furthermore, it represents quite good an approximation also for the
three-dimensional frozen-planet configuration, since the latter is
characterized by a very small overlap between the outer and the inner
electron's densities, and therefore by a very small splitting between
singlet and triplet states \cite{RicO92JPB}.
Precisely as for the classical Hamiltonian, the external electric
field consists of an oscillating part $F$ -- which, in contrast to
(\ref{hcl}), is incorporated in the velocity gauge in order to ensure
better convergence of the numerical calculation \cite{Sha88ZPD} -- and
of a static component $F_{st} > 0$ which forces the electrons away
from the nucleus.

In a last step, we introduce the coordinates 
\begin{equation}
\xi = z_1 - z_2\, , \eta = z_2\, ,
\label{pekeris}
\end{equation}
which independently lie in the range $0 \leq \xi, \eta
\leq \infty$.
In these {\em perimetric coordinates} \cite{Pek58PR}, the Hamiltonian is
rewritten as 
\begin{eqnarray}
H & = & - \frac{\partial^2}{\partial \xi^2} 
  \; - \; \frac{1}{2} \frac{\partial}{\partial \eta^2} 
  \; + \; \frac{\partial^2}{\partial \xi \partial \eta}
  \; - \; \frac{Z}{\xi + \eta} \; - \; \frac{Z}{\eta} 
  \; + \; \frac{1}{\xi} \nonumber \\
&& \; - \; \frac{F}{\omega} \, \sin \omega t \, \frac{1}{\rm i} 
  \frac{\partial}{\partial \eta} \; - \; F_{st} \, ( \xi + 2 \eta ) ,
  \label{Hper}
\end{eqnarray}
what will turn out as a convenient form for our further treatment.

\subsection{Floquet theory and complex scaling}

Due to the temporal periodicity of the Hamiltonian, the Schr{\"o}dinger
problem represented by (\ref{Hper}) is conveniently treated in the framework
of 
Floquet theory \cite{Shi65PR,Zel67JETP}:
Any solution of the Schr{\"o}dinger equation can be 
expanded in a set of $2\pi/\omega$-periodic 
quasienergy-eigenfunctions $\psi_t^{({\mathcal E})}$,
\begin{equation}
\psi_t = \int\hspace*{-0.5cm}\sum d {\mathcal E}  C_{\mathcal
  E} \psi_t^{({\mathcal E})} e^{ - {\rm i} {\mathcal E} \, t
  }, \, \mbox{with} \, \psi_{t + 2 \pi / \omega}^{({\mathcal
  E})} = \psi_t^{({\mathcal E})}\, , C_{\mathcal E}\in{\mathbb C}
\end{equation}
which satisfy the Floquet eigenvalue equation
\begin{equation}
\left( H - {\rm i} \, \frac{\partial}{\partial t} \right) 
\psi_t^{({\mathcal E})} \; = \; {\mathcal E} \, \psi_t^{({\mathcal E})}.
\label{genee}
\end{equation}
A Fourier series expansion of $\psi_t^{({\mathcal E})}$,
\begin{equation}
\psi_t^{({\mathcal E})} = \sum_{k=-\infty}^{\infty}
  \hat{\psi}_k^{({\mathcal E})} e^{{\rm i} \, k \, \omega \, t},\label{fourier}
\end{equation}
yields a {\em time-independent} eigenvalue problem for the
Fourier components $\hat{\psi}_k^{({\mathcal E})}$, where the effective
Hamiltonian matrix contains the stationary part of the Hamiltonian
shifted by $k \omega$ as diagonal, and the Fourier
components of the periodic driving as off-diagonal (block) elements.
This results in an $\omega$-periodic spectrum of quasienergies
${\mathcal E}$.

For atomic systems, the Floquet spectrum is absolutely continuous:
each bound state of the unperturbed atom is coupled to the atomic 
continuum via multiphoton transitions, and therefore appears as a
resonance structure in the spectrum  -- in analogy to autoionizing
states in multiply excited atoms, which are coupled to the continuum
via configuration interaction \cite{Fan61PR} (see, e.g.,
\cite{MadSchLam00PRL} for a case in triply excited lithium).
In order to separate these resonances from the flat background of the
continuous spectrum, we use the method of complex scaling 
\cite{BalCom71CMP,Ho83PSR,Rei82ARPC}. 
It consists in the complexification of coordinates and momenta ({\em not} of
time!) according to \footnote{Note that, due to (\ref{pekeris}),
complexification of $\xi,\eta$ and of $z_1,z_2$ is equivalent.}
\begin{eqnarray}
\xi \; \longrightarrow \; \xi e^{i\theta} & \qquad & -i\frac{\partial}{\partial\xi} \; \longrightarrow \; -i\frac{\partial}{\partial\xi} e^{-i\theta} 
\nonumber \\
\eta \; \longrightarrow \; \eta e^{i\theta} & \qquad & -i\frac{\partial}{\partial\eta} \; \longrightarrow \; -i\frac{\partial}{\partial\eta} e^{-i\theta},
\label{scale0}
\end{eqnarray}
through application of the {\em nonunitary} complex scaling operator
$R(\theta)$ on $\psi_t^{({\mathcal E})}$,
\begin{equation}
R(\theta)\psi_t^{({\mathcal E})}(\xi,\eta)=e^{i\theta}\psi_t^{({\mathcal E})}(\xi e^{i\theta},\eta
e^{i\theta})=:\psi_{\theta ,t}^{({\mathcal E})}(\xi,\eta)\, .
\label{scaleop}
\end{equation}
Accordingly, the Hamiltonian is transformed as 
\begin{equation}
H \; \longrightarrow \; H_{\theta}=R ( \theta ) \, H \, R ( - \theta ),
\label{scale1}
\end{equation}
what turns (\ref{genee}) into a complex symmetric rather than hermitian
eigenvalue problem, with eigenvalues in the lower half of the complex plane.
In particular, resonances of the
``real'' eigenvalue problem (i.e., at $\theta = 0$) appear, for finite 
$\theta$, as isolated, {\em discrete} complex eigenvalues
${\mathcal E} = E - {\rm i} \Gamma/2$: their real and imaginary parts 
correspond to the energies $E$ and to the half widths at
half maximum $\Gamma/2$ (HWHM) of the resonances, respectively.

\subsection{Expansion in Sturmian basis functions}

The complex-scaled Floquet Hamiltonian is now expanded in the product
basis
\begin{equation}
\{ S_n^{(\alpha)}(\xi) S_m^{(\beta)}(\eta)\, : n,m \geq 1 \} \label{stbasis}
\end{equation}
composed of the real-valued Sturmian functions \cite{BluSmi87ZPD}
\begin{equation}
S_n^{(\alpha)} ( \xi ) \; = \; \frac{(-1)^n}{\sqrt{n}} \, \frac{2 \xi}{\alpha }
  \,  
  \exp \left( - \frac{\xi }{\alpha } \right) L_{n - 1}^{(1)} \left( 
  \frac{2 \xi }{\alpha } \right),
\label{st}
\end{equation}
where the $L_{n-1}^{(1)}$ denote the associated Laguerre polynomials
\cite{AbrSte}. 
The real parameter $\alpha > 0$ fixes the length scale of 
the Sturmians.
As a consequence, the product basis (\ref{stbasis}) exhibits two
independent scaling parameters, $\alpha$ and $\beta$, which permit the
optimal tuning of the basis with respect to the relative excitation
of the 
electrons.
Also note 
that the Sturmians 
$S_n^{(\alpha )}$ scale at
least linearly with $\xi $, for $\xi \to 0$.
This implies that the Coulomb singularities of the Hamiltonian
(\ref{Hper}) do not lead to divergent matrix elements in the
product basis (\ref{stbasis}).

Multiplication of (\ref{genee}) by $\xi\eta$ leaves $1 / ( \xi + \eta )$ as
the only non-polynomial contribution in $\xi$ and $\eta$, after inserting
(\ref{Hper}). We factor out $( \xi + \eta )$ from the eigenvectors of
(\ref{genee}), 
\begin{equation}
\psi_t^{({\mathcal E})} ( \xi, \eta ) \; =: \; ( \xi + \eta ) \,
\phi_t^{({\mathcal E})} ( \xi, \eta )\, ,
\label{phit}
\end{equation}
and expand $\phi_t^{({\mathcal E})} ( \xi, \eta )$ in the Sturmian
product basis.
This factorization is justified since the triple collision between the
electrons and the nucleus -- i.e., the event that corresponds to 
$\xi = \eta = 0$ -- is classically suppressed in the collinear frozen-planet
dynamics (in contrast to the $eZe$ configuration with the electrons on
opposite sides of the nucleus);
hence, the quantum probability near $\xi = \eta = 0$ is very small.
Note that, due to normalization, (15) entails a different transformation law
for $\phi_t^{({\mathcal E})}(\xi, \eta)$ as compared to 
$\psi_t^{({\mathcal E})}(\xi, \eta)$, under the action of $R(\theta)$:
\begin{equation}
R(\theta)\phi_t^{({\mathcal E})}(\xi, \eta)=e^{2i\theta}\phi_t^{({\mathcal
    E})} (\xi e^{i\theta}, \eta e^{i\theta})  =: 
\phi_{\theta , t}^{({\mathcal E})}(\xi, \eta)\, .
\end{equation}

Alltogether, the 
generalized 
Floquet eigenvalue
equations finally
reads
\begin{eqnarray}
& &   ( \xi + \eta ) \xi \eta ( H_{0 \theta} + k \omega -
  {\mathcal E} ) ( \xi + \eta ) \hat{\phi}_{\theta, k}^{({\mathcal E})} ( \xi,
  \eta ) +  ( \xi + \eta ) \nonumber \\
& &   \times  \xi  \eta  V_{\theta}  ( \xi + \eta )
   ( \hat{\phi}_{\theta ,k - 1}^{({\mathcal E})} ( \xi, \eta ) - 
\hat{\phi}_{\theta ,k +
  1}^{({\mathcal E})} ( \xi, \eta ) )  =  0\, ,
\label{evscale}
\end{eqnarray}
with 
the complex
scaled operators $H_{0 \theta}$, $V_{\theta}$ 
given by
\begin{eqnarray}
H_{0 \theta} & = & \left( - \frac{\partial}{\partial \xi^2} -
  \frac{1}{2} \frac{\partial}{\partial \eta^2} +
  \frac{\partial^2}{\partial \xi \partial \eta} \right) {\rm e}^{- 2
  {\rm i} \theta} \nonumber \\ 
&& + \left( - \frac{Z}{\xi + \eta} - \frac{Z}{\eta} + \frac{1}{\xi} 
  \right) {\rm e}^{- {\rm i} \theta} \nonumber \\  
&& -  F_{st} \, ( \xi + 2 \eta ) 
  \, {\rm e}^{{\rm i} \theta}, \label{Hflc} \\
V_{\theta} & = & \frac{F}{2 \omega} \, \frac{\partial}{\partial \eta}
  \, {\rm e}^{- {\rm i} \theta}, \label{Dflc}
\end{eqnarray}
and the $\hat{\phi}_{\theta, k}^{({\mathcal E})}(\xi,\eta)$ 
the Fourier
components (see (\ref{fourier}))
of the wavefunction $\phi_{\theta,t}^{(\mathcal E)}(\xi,\eta)$. 
Introducing the scalar product
\begin{equation}
\widetilde{\langle f \, | g \rangle} \; \equiv \; \int_0^{\infty} 
\int_0^{\infty} \frac{1}{\xi \, \eta}
\, f ( \xi, \eta ) \, g ( \xi, \eta ) \, d \xi \, d \eta \, , \label{sp}
\end{equation}
with respect to which the Sturmians are orthogonal,
we obtain a complex symmetric eigenvalue problem.

All the opertor-valued terms acting on the $\hat{\phi}_{\theta, k}^{({\mathcal
E})}$ and $\hat{\phi}_{\theta, k\pm1}^{({\mathcal E})}$ in (\ref{evscale}) 
can now 
be expressed as polynomials of the elementary ladder operators of the
Sturmian basis (\ref{st}) (see \cite{Sch99,AbrSte}).
Hence, upon expansion of the Fourier components
$\hat{\phi}_{\theta, k}^{({\mathcal E})}$ in the Sturmian product basis
(\ref{stbasis}), 
\begin{equation}
  \hat{\phi}_{\theta, k}^{({\mathcal E})} ( \xi, \eta ) \; = \; \sum_{m, n =
  1}^{\infty} C_{n,m}^{(k)} \, S_{n}^{(\alpha)} ( \xi ) \,
  S_{m}^{(\beta)} ( \eta ) \label{phik_st}\, ,
\end{equation}
we obtain strict selection rules which permit nonzero couplings
between the coefficients $C_{n,m}^{(k)}$ and $C_{n',m'}^{(k')}$ only
if $|n - n'| \leq 4$, $|m - m'| \leq 4$, and $|k - k'| \leq 1$.
The eigenvalue problem thus becomes a sparse, complex symmetric
matrix equation with rather narrow bandwidth, what allows to apply efficient
diagonalization methods based on the Lanczos algorithm with inverse
iteration \cite{DelBomGay91PRL}. 
Convergence of the complex eigenvalues is contolled by increasing the
number of Fourier components $\hat{\phi}_{\theta, k}^{({\mathcal E})}$ 
and of the basis functions $S_n^{(\alpha )}(\xi )$, $S_m^{(\beta )}(\eta )$
to be retained
in the diagonalization, as well as by variation of the complex scaling
angle $\theta$, and of the length scale parameters $\alpha$ and $\beta$
which specify the (truncated) basis set (\ref{st}) used in the
calculation.
All numerical results presented hereafter are converged to
machine precision.

\subsection{Visualization of the wavefunction}

\label{vswf}

The energies $E$ and widths $\Gamma $ extracted from the complex
eigenvalues ${\mathcal E}$ allow to classify the Floquet states in
series and to determine their stability; in general, however, they do
not provide any direct information about the localization properties of these
states in configuration or in phase space, which in themselves can carry
important physical information. 
To extract the latter from the eigenstates of the complex symmetric eigenvalue
problem (\ref{evscale}), some care has to be taken, due to the non-unitarity of
the dilation operator $R(\theta )$. The technical and computational details of
this procedure are described in detail in \cite{Sch99,BucGreDel94JPB}, and we
only import here those results of immediate relevance for our specific
purpose. 

We first need the density of the physical wave function 
(at {\em real} energy $E$)
$\psi_t^{({\mathcal E})}(z_1,z_2)$, which is obtained (up to a normalization
constant), for a well-isolated
resonance energy ${\mathcal E}=E-i\Gamma/2$ (i.e., $|{\mathcal E}_j-{\mathcal
E} | \gg\Gamma\, ,\forall {\mathcal E}_j\neq {\mathcal E}$), from the image of  
$\psi_{\theta ,t}^{({\mathcal E})}$ under the inverse dilation operator
$R(-\theta )$, 
\begin{equation}
  R(-\theta )\psi_{\theta ,t}^{({\mathcal E})}=  
R(-\theta) \sum_{k = - \infty}^{\infty} z_1 \hat{\phi}_{\theta, k}^{({\mathcal E})} 
\left( z_1 - z_2, z_2 \right) {\rm e}^{ {\rm i} k \omega t }\, , \label{rerot}
\end{equation}
together with the image of the corresponding left eigenvector $\psi_{\theta
,-t}^{({\mathcal E})}$ \cite{Sch99,BucGreDel94JPB}:\footnote{Only the
time-dependent part of the wave  
function is complex conjugated to transform right into left eigenvectors, 
since the Fourier components 
$\hat{\psi}_{\theta, k}^{({\mathcal E})}$ of the
eigenvectors obey the normalization condition for complex symmetric rather
than hermitian matrices
\protect\cite{Buc93,Sch99}!}
\begin{equation}
|\psi_t^{({\mathcal E})}(z_1,z_2)|^2\simeq {\rm Re}\langle
 z_1,z_2|R(-\theta)|\psi_{\theta, t}^{({\mathcal E})}\rangle
 \langle z_1,z_2|R(-\theta)|\psi_{\theta, -t}^{({\mathcal E})}\rangle  \, .
\label{confi}
\end{equation}
Correspondingly, the Husimi representation $Q(z_1,p_1)$
\cite{Sch99,BucGreDel94JPB} in the classical phase space coordinates of the
{\em outer} electron (subscript 1) is obtained by projection on Gaussian wave
packets $\chi_{p_1}(z_1)$ which are centered at position $z_1$ and propagate
with average momentum $p_1$: 
\begin{equation}
Q(z_1,p_1)\simeq {\rm Re} \langle\chi_{p_1}(z_1) | R(-\theta)| \psi_{\theta
  ,t}^{({\mathcal E})}\rangle 
  \langle \overline{\chi_{p_1}(z_1)} | R(-\theta)|\psi_{\theta, -t}^{({\mathcal
  E})}\rangle \, ,
\label{hus}
\end{equation}
where the bar denotes complex conjugation.

In order to establish a precise analogy with our above construction of the
two-step Poincar{\'e} section of the classical dynamcis (see Section \ref{PC}),
we define the overlap matrix element in (\ref{hus}) as 
\begin{eqnarray}
&&  \langle \chi_{p_1}(z_1) | R(-\theta)|\psi_{\theta, t}^{({\mathcal E})}
  \rangle =  
\int_0^{\infty} 
  d z_1' \, \langle z_1',z_2^{(0)}|R(-\theta)|\psi_{\theta, t}^{({\mathcal
E})}\rangle \nonumber \\ 
&& \times   \exp \left( - \frac{1}{2}  
    \omega_s (z_1 - z_1')^2 - {\rm i} p_1 z_1' \right). \label{husimi}
\end{eqnarray}
Thus, the 
quantum phase space probability density is evaluated at
fixed time $t$ (corresponding to a fixed phase $\omega t$ of the driving
field), and at fixed position $z_2^{(0)} \simeq 0$ of the inner electron.
The {\em squeezing parameter} $\omega_s$ in this definition determines the
resolution of the Husimi distribution in position and momentum,
respectively.
Large values of $\omega_s$ yield good resolution in position space and
bad resolution in momentum space, and vice versa for small values of
$\omega_s$.
For quantum states associated with the frozen-planet configuration,
the most appropriate choice for $\omega_s$ is given by the intrinsic
frequency scale $\omega_I$ of the configuration (see Section \ref{FP})
-- this is the value at which the Gaussian wave packet $\chi_p(q)$
optimally fits 
the ground state of the effective potential
experienced by the outer electron.
We therefore choose $\omega_s = 0.3 \, N^{-3}$ (see Fig.~\ref{adiab}), 
with $N$ the inner 
electron's principal quantum number of the state to be visualized
(note that $N$ remains approximately a good quantum number in the
presence of the external field, if the latter is not too strong).

\section{The unperturbed collinear atom}

\label{1D}

With the above theoretical machinery for our quantum treatment of the driven,
collinear frozen planet configuration at hand, we can now study in detail its
physical properties. As a ``warm-up'', and equally so as a first assessment of
the physical implications of the restricted dimensionality of our problem, we
start out with the spectral properties of the {\em unperturbed} ($F=F_{\rm
st}=0$) frozen planet configuration confined to a single dimension of
configuration space. 

\subsection{Energies and phase space distributions}

\begin{figure*}
\begin{center}
\leavevmode
\epsfxsize15cm
\epsfbox{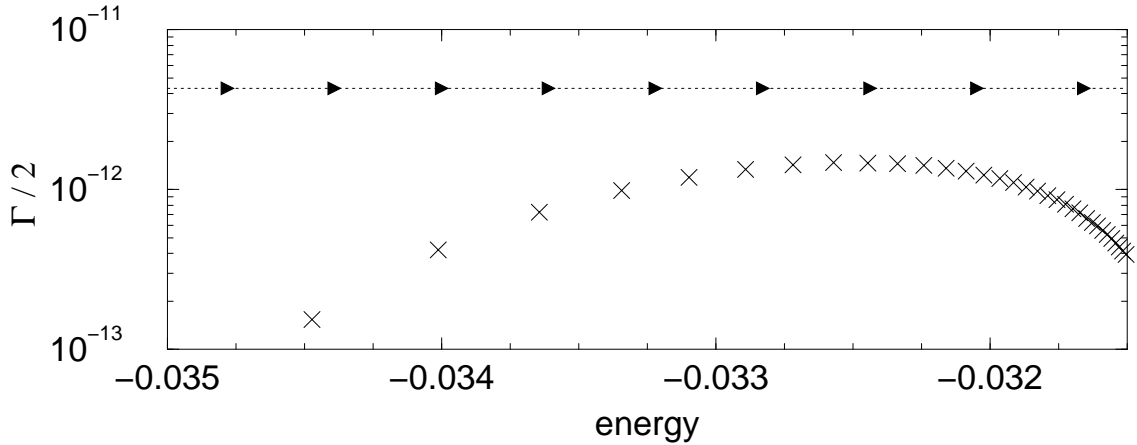}
\end{center}
\caption{Energies and ionization rates $\Gamma / 2$ (HWHM) of the 
  autoionizing states of collinear helium 
belonging to the series labeled by the
  inner electron's quantum number $N = 8$. 
  Note the relatively small widths of the lowest members of the series,
  which we attribute to the fact that these states are fully localized 
  within a regular, bound domain in phase space.
  The coupling mediated by a driving field with frequency $\omega =
  0.2 \, N^{-3}$ is sketched by the arrows of length $\omega$.
  The particularly strong, near-resonant coupling between the second and 
  the third state (at $E = -0.034$ and $E = -0.0336$, respectively) gives
  rise to the nondispersive two-electron wave packet associated with the 
  classical 1:1 resonance (see Fig.~\ref{ndwp}).
}
\label{spn8}
\end{figure*}

Corresponding to the fact that its classical counterpart is almost 
completely integrable (Fig.~\ref{pspF}(a)), the quantum spectrum of the 
unperturbed collinear frozen-planet configuration is characterized by 
typical signatures of regular dynamics.
In particular, each autoionizing state can be classified by two quantum
numbers which are associated with the two separating modes of the
classical dynamics, as discussed in Section \ref{FP} (in contrast to
the chaotic $eZe$ configuration \cite{BluRei91Col}, where such a
classification breaks down already for moderate double excitations).
The quantization of the inner electron's Kepler mode gives rise to the
quantum number $N \geq 1$.
For each $N$, we obtain a series of autoionizing states (or bound states,
for $N = 1$) which are labelled by the quantum number $n$.
This quantum number arises from the quantization of the outer electron's
oscillation within the effective potential, which, as in one-electron atoms,
gives rise to Rydberg series and continua, due to its asymptotic $1/r$
dependence.

As an example, we show the energies and widths of the
autoionizing states of the series labeled by $N = 8$ in Fig.~\ref{spn8}.
We clearly recognize a regular Rydberg progression towards the $N=8$
threshold at $E = - 0.3125$ a.u. 
States belonging to different series are energetically well separated:
The lowest quantum state with $N = 9$ lies at $E \simeq -0.027$ a.u., and 
the Rydberg series of $N = 7$ terminates at $E \simeq - 0.04$ a.u.
As was verified numerically, an energetic overlap of series belonging
to different quantum numbers $N$ occurs only above $N = 20$.

\begin{figure*}
\begin{center}
\leavevmode
\epsfxsize15cm
\epsfbox{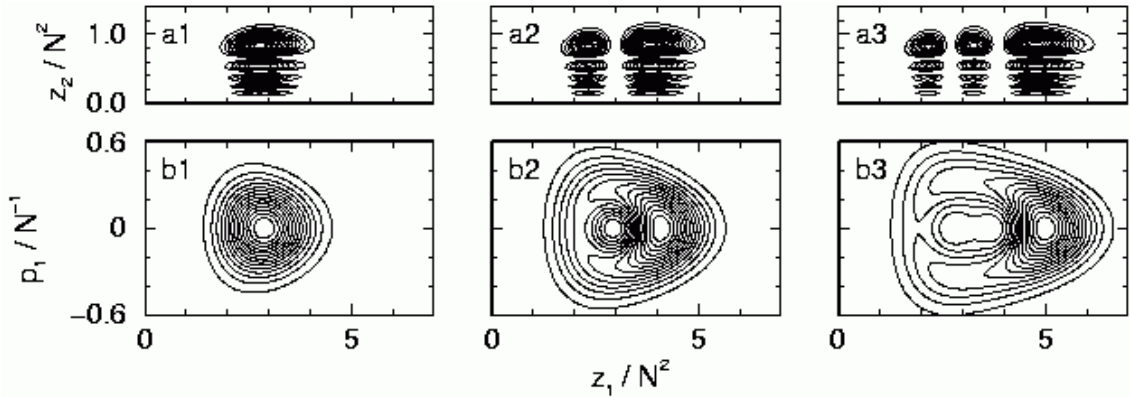}
\end{center}
\caption{Wavefunctions of the first three eigenstates of the $N = 8$
  series: (a1, b1) lowest ($n = 1$), (a2, b2) second ($n = 2$), (a3,
  b3) third state ($n = 3$) of the series.
  (a1 -- a3) Probability densities as a function of the coordinates
  of the outer ($z_1$) and of the inner ($z_2$) electron;
  (b1 -- b3) Husimi densities as a function of position and momentum of
  the outer electron (the contour lines are plotted on a linear scale).
  In the coordinate of the inner electron, the wavefunctions are
  essentially given by the $N$th hydrogenic eigenstate of the
  unperturbed, one-dimensional atom, whereas in $z_1$ they display
  different excitations within the effective potential (see
  Fig.~\ref{adiab}).
  Their Husimi distributions clearly show that they are well localized 
  along regular tori in the underlying classical phase space (see 
  Fig.~\protect{\ref{pspF}}(a)).
  \label{dshs}}
\end{figure*}

The near-separability of the two modes associated with the electrons
is furthermore reflected by the probability density (\ref{confi}) of the
eigenstates (with $k=0$ in (\ref{rerot}), due to the time-independence of the
problem).
In the coordinate $z_2$ of the inner electron, the wavefunctions are
essentially given by the $N$th hydrogenic eigenstate of the
one-dimensional atom (most reminiscent of extremal parabolic states
of three-dimensional atomic hydrogen), whereas in $z_1$ they
display distinct excitations within the effective potential governing
the outer electron's motion.
In Fig.~\ref{dshs}, the fundamental (a1), the first excited (a2), and
the second excited state (a3) of the effective potential (for $N =
8$), labeled by the outer electron's quantum numbers $n = 1$, $n 
= 2$, and $n = 3$, respectively, are plotted in configuration space.
The intimate correspondence of these eigenstates with regular
structures of the underlying classical phase space is highlighted 
by their Husimi distribution (\ref{hus}) 
shown in Fig.~\ref{dshs}(b1 -- b3).
Comparing these phase space projections of the quantum eigenfunctions
with the corresponding Poincar\'{e} surface of section 
(Fig.~\ref{pspF}(a)) unambiguously illustrates the neat localization
of the eigenfunctions along invariant tori of the classical dynamics.

\subsection{Ionization rates}

\label{1Dw}

\begin{figure*}
\begin{center}
\leavevmode
\epsfxsize15cm
\epsfbox{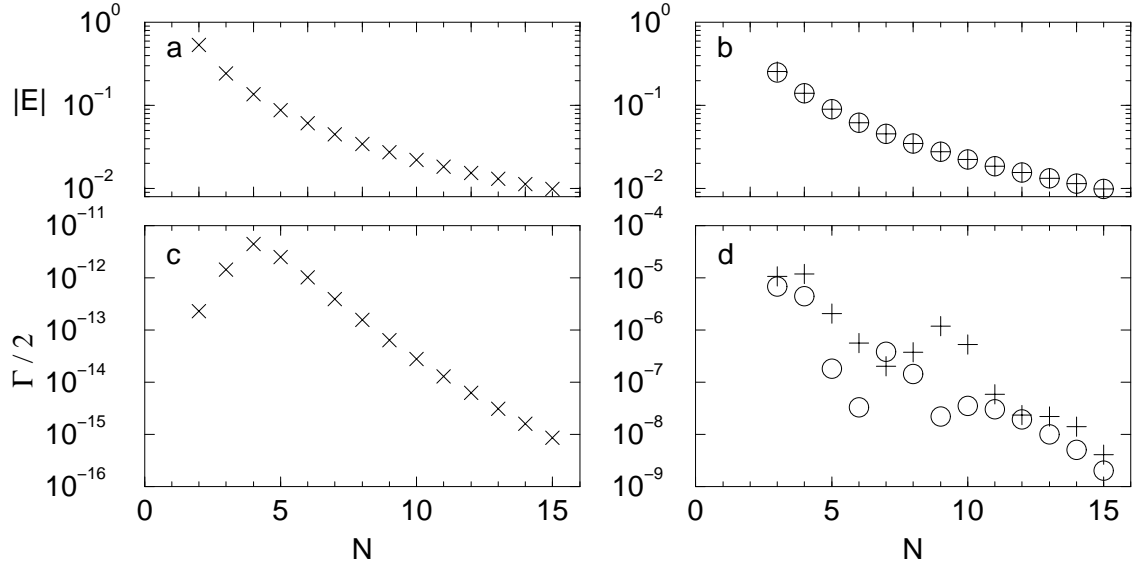}
\end{center}
\caption{Energies (a) and ionization widths (c) of the lowest 
  autoionizing eigenstate $n = 1$ (with the outer electron localized
  at the minimum of the effective potential of Fig.~\ref{adiab}),
  plotted as a function of the inner electron's quantum number $N$.
  We see that the ionization rates decrease near-exponentially for
  $N > 4$, what suggests that the associated eigenstates are fully 
  localized in the bound part of phase space, and only decay via tunneling 
  through the classical phase space barriers that protect the
  frozen-planet configuration against classical autoionization.
  These rates of the collinear model 
are compared to the energies (b) and decay rates
  (d) of the analogous frozen-planet eigenstates in
  three-dimensional helium, published in Ref.~\protect\cite{RicO92JPB}, for
  singlet ($+$) and triplet ($\bigcirc$) symmetry.
  While the energies agree quite well with their collinear counterparts
  (a, b), the ionization rates of the three-dimensional frozen-planet 
  states (d) lie by several orders of magnitude above the widths of the 
  corresponding eigenstates of the one-dimensional model (b) (note the
  different scales of the vertical axes in (b) and (d)).
  This indicates that the autoionization of the three-dimensional 
  frozen-planet states is mediated by the transverse degrees of freedom
  of the configuration. \label{enwd}}
\end{figure*}

The ``regular'' character of the wavefunctions is also manifest in the
ionization rates $\Gamma / 2$ 
(HWHM) of the associated spectral resonances.
We see in Fig.~\ref{spn8} that for the $N = 8$ series these widths
first increase with $n$, and then decrease again towards the ionization
threshold.
While this latter decrease is attributed to general scaling laws of $\Gamma$
in the Rydberg regime of atomic one-electron excitations \cite{Fri}, the
extraordinary stability of low $n$ states is a consequence of their
full localization in the bound part of phase space.
A coupling to the unbound, ionizing domain is then only possible via a
classically forbidden process -- tunneling through the phase space
barriers that confine the frozen-planet configuration.

This latter mechanism is beautifully illustrated in a 
plot of the ionization
rate of the lowest state ($n = 1$) as a function of the inner
electron's quantum number $N$. 
In Fig.~\ref{enwd}(c), we observe that the width of this lowest
eigenstate in the effective potential of Fig.~\ref{adiab} first
increases for $N \leq 4$ (due to imperfect localization of the
eigenstate on classical phase space structures in the deep quantum
regime) and then decreases exponentially with $N$.
This exponential law represents a clear signature of quantum tunneling:
The rate associated with such a process generally decreases exponentially 
with the semiclassical action $S_N$ across the tunneling barrier, and
this latter quantity scales, as any other action variable, linearly
with $N$, due to the general scaling laws (\ref{scr}) --
i.e., $S_N = N S_1$, with $S_1 \simeq 0.8$ directly extracted from
Fig.~\ref{enwd}(c).

For comparison, we plot in Fig.~\ref{enwd}(d) the ionization rates of
the near-collinear frozen-planet states in three-dimensional helium
(with minimal excitations in the transverse degrees of freedom), as
reported in \cite{RicO92JPB}.
Also these widths decrease on average exponentially in $N$, with a
scaled action $S_1$ not too different from our one-dimensional model
(fluctuations with respect to this exponential decrease arise from the
mixed regular-chaotic phase space structure of three-dimensional
helium, see also the discussion in Section \ref{LW}). 
However, the three-dimensional rates are systematically enhanced by
several orders of magnitude as compared to the one-dimensional model!
This suggests that the three-dimensional helium dynamics provides
efficient decay channels associated with the {\em transverse} 
degrees of freedom, which are not incorporated in the restricted
one-dimensional dynamics of our model Hamiltonian (\ref{Hqm}).\footnote{As a
side remark, this observation also bears a {\em caveat} as oversimplified
models of the three body Coulomb problem are concerned: often the Coulomb
singularities in such 1D model systems are smoothed such as to prevent the
electrons (and possibly the nucleus) from head-on collisions
\cite{LapO96JPB,RzaLewSal94PRA}, which, on a first glance, could be most
desastrous
for the stability of the 1D dynamics and, hence, {\em enhance} its decay
probability. The present comparison of exact (no approximations beyond
dimensionality) 1D and 3D quantum calculations shows that this is too
simplistic an argument and completely neglects dynamical stabilization effects
which result in such impressively stable quantum objects as the frozen
planet.}  

On the other hand, the {\em energies} of the autoionizing states of
three-dimensional helium agree quite well with their collinear 
counterparts, as apparent from Figs.~\ref{enwd}(a) and \ref{enwd}(b)
(apart from a small shift due to the additional quantization of the
transverse modes in the 3D configuration \cite{RicO92JPB}).
This supports our expectation that the essential structure of the
eigenstates of the three-dimensional frozen-planet configuration is
well reproduced by the collinear model.

\section{The collinear atom under periodic driving}

\label{WP}

\subsection{Nondispersive wave packets in the Floquet spectrum}

\label{NW}

For our investigation of the collinear frozen planet configuration
under external driving, we employ the field parameters $\omega = 0.2
\, N^{-3}$ and $F = 0.005 \, N^{-4}$ which induce the classical phase
space structure shown in Fig.~\ref{psp_rot}, where both the intrinsic
island and the 1:1 resonance are very well pronounced.
The coupling induced by this driving field is illustrated by the
arrows in Fig.~\ref{spn8} for the $N = 8$ series.
We see that a particularly strong interaction is induced between the
second ($n = 2$) and the third ($n = 3$) state, since the energetic
spacing between these states is very close to the driving frequency.
Provided the phase space volume of the 1:1 resonance island is large
enough to support a fully localized quantum state (which is indeed the
case for $N=8$, as can be semiclassically estimated by the EBK
quantization criterion), this near-resonant one-photon coupling gives
rise to the nondispersive wave packet associated with the classical
nonlinear resonance.

\begin{figure*}
\begin{center}
\leavevmode
\epsfxsize15cm
\epsfbox{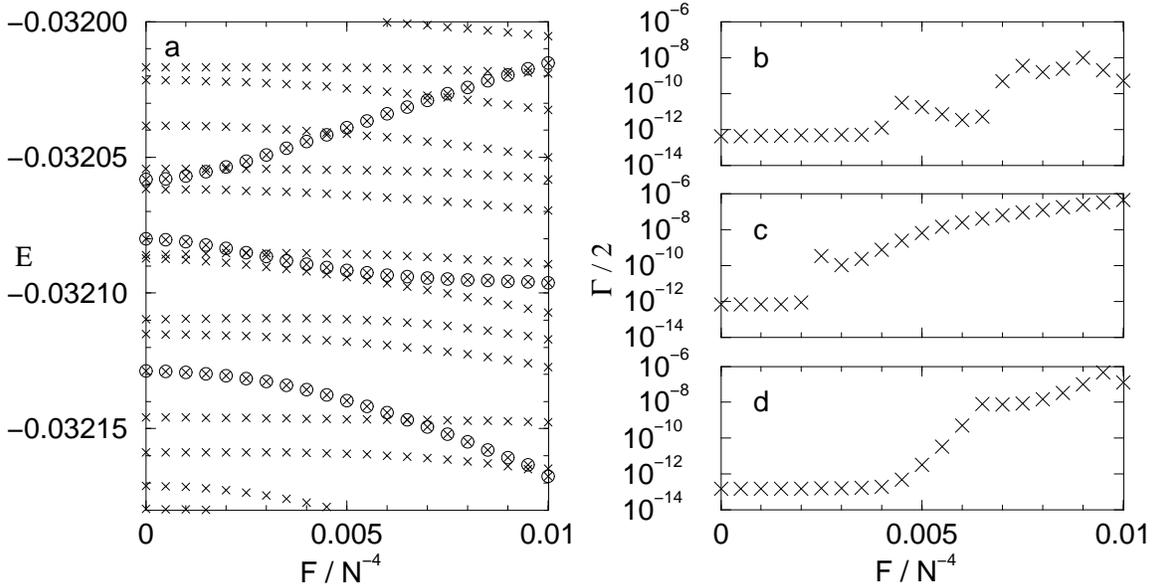}
\end{center}
\caption{(a) Evolution of the first three states of the $N = 8$ series
  in the Floquet spectrum, for driving frequency $\omega = 0.2 \,
  N^{-3}$.
  The quasienergies $E$ are plotted as a function of the field
  amplitude $F$.
  The diabatic continuation of the second state ($n = 2$, upmost
  circles) undergoes a particularly pronounced, almost linear
  shift with $F$.
  This is consistent with the wave packet character of the
  Floquet state into which the unperturbed $n = 2$ state evolves (see
  Fig.~\ref{ndwp}).
  (b -- d) Ionization rates $\Gamma$ of the diabatic continuations of the 
  $n = 1$ (d), $n = 2$ (b), and $n = 3$ state (c).
  We see that the rates globally increase with increasing $F$.
  Local, pronounced maxima occur at avoided crossings between these states
  and energetically higher-lying eigenstates of the $N = 8$ series in
  the Floquet spectrum, as apparent from a comparison with (a) (with the
  diabatic continuations of $n = 1$ and $n = 3$ marked by the encircled
  eigenvalues in the lower and the middle part of (a), respectively).
\label{flsp}}
\end{figure*}

Fig.~\ref{flsp}(a) shows the evolution of the quasienergies associated
with the first three Floquet eigenstates of the $N = 8$ series, as a
function of the field amplitude $F$ (for fixed driving frequency
$\omega = 0.2 \, N^{-3}$, and $F_{st} = 0$).
In the limit $F = 0$, the energies $E$ correspond to the unperturbed
autoionizing levels folded into the Floquet zone due to the
$\omega$-periodicity of the Floquet spectrum.
Switching on the driving field causes level shifts due to the
repulsion between strongly coupled, near-resonant Floquet states.
A particularly significant case is encountered for the state $n = 2$,
near-resonantly coupled to the states $n = 3$ and $n = 1$.
Its quasienergy exhibits a pronounced shift at almost constant, positive slope.
This already indicates the wave-packet character of the associated
eigenfunction in configuration space, since it reflects a large
dynamic dipole moment of the electronic density 
(proportional to the expectation value of the electronic
dipole $z_1$, averaged over one cycle of the drive).

\begin{figure*}
\begin{center}
\leavevmode
\epsfxsize15cm
\epsfbox{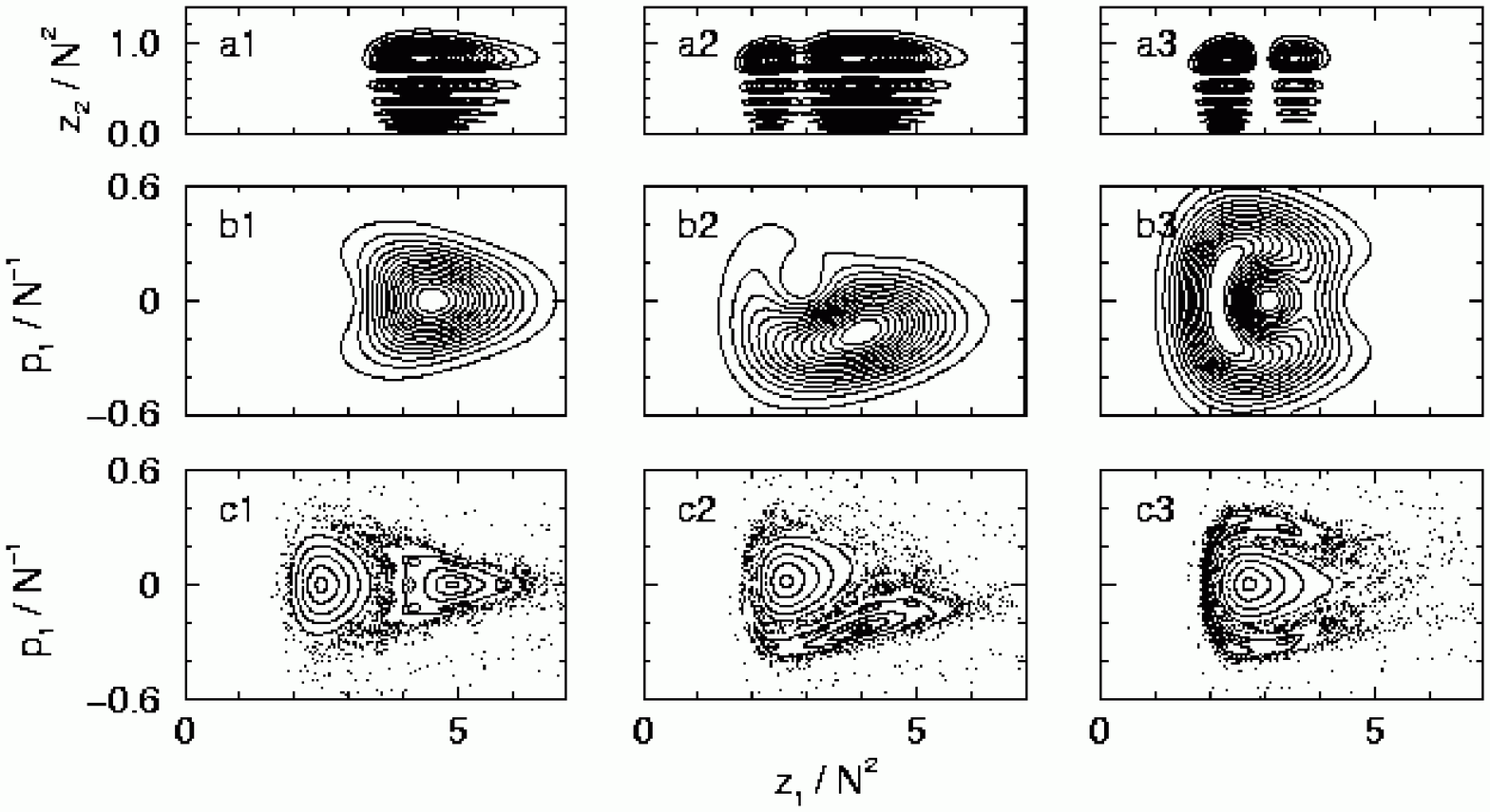}
\end{center}
\caption{Nondispersive two-electron wave packet of the $N = 8$ series,
  for $\omega = 0.2 \, N^{-3}$ and $F = 0.005 \, N^{-4}$.
  We plot the probability density in configuration space (a1 -- a3),
  and the Husimi density in phase space (b1 -- b3), of the diabatic
  continuation of the state $n = 2$ (see Fig.~\ref{flsp}), at driving
  field phases (a1 -- c1) $\omega t = 0$, (a2 -- c2) $\pi /
  2$, (a3 -- c3) $\pi$.
  The corresponding Poincar\'{e} sections of the classical phase space
  are plotted in (c1 -- c3).
  We see that the Floquet wavefunction is well localized on the 1:1 
  resonance island, and faithfully tracks its classical time
  evolution.
\label{ndwp}}
\end{figure*}

The wave packet character of this latter state is ultimately confirmed
by its Husimi density, which is plotted in Fig.~\ref{ndwp}(b1 -- b3)
for $F = 0.005 \, N^{-4}$, at driving phases $\omega t = 0$, $\pi / 2$,
and $\pi$.
We see that the Floquet eigenfunction is fully localized on the 1:1
resonance island in phase space, and that it precisely follows the
classical time evolution.
The wave packet dynamics in configuration space is visualized in 
Fig.~\ref{ndwp}(a1 -- a3) where we plot the probability density of the
state $n = 2$ as a function of $z_1$ and $z_2$.
Comparison with Figs.~\ref{dshs}(a1 -- a3) shows that the
eigenfunction is essentially composed of a time-periodic superposition
of the low-lying eigenstates of the $N = 8$ series.
While nearly stationary in the coordinate $z_2$ of the inner electron,
the wave packet oscillates back and forth in $z_1$, between the
turning points of the effective potential (Fig.~\ref{adiab}).

\begin{figure*}
\begin{center}
\leavevmode
\epsfxsize15cm
\epsfbox{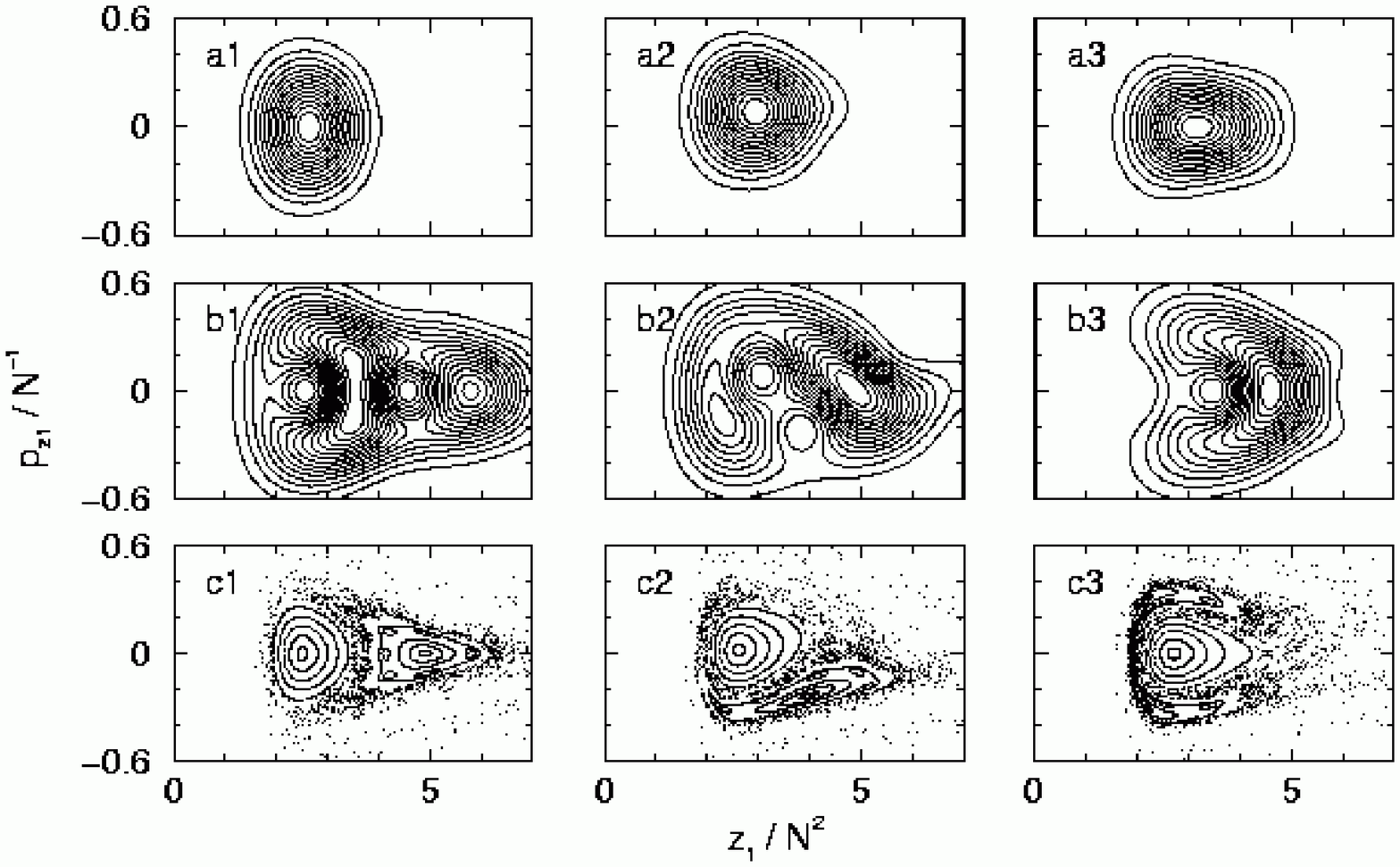}
\end{center}
\caption{Husimi distribution of the states $n = 1$ (a1 -- a3) and 
  $n = 3$ (b1 -- b3) of the $N = 8$ series, for $\omega = 0.2 \,
  N^{-3}$ and $F = 0.005 \, N^{-4}$, at driving phases $\omega t = 0$
  (a1,b1), $\pi / 2$ (a2,c2), $\pi$ (a3,b3), compared 
to the corresponding classical Poincar\'{e} sections
  (c1 -- c3). 
  We see that the diabatic continuation of the lowest state of the $N
  = 8$ series is fully localized on the intrinsic island and remains
  essentially stationary over one field cycle.
  The diabatic continuation of the third state, on the other hand, is
  associated with the chaotic phase space domain.
  More precisely, it is anchored to the unstable fixed point of the
  1:1 resonance and stretches out along the associated separatrix
  structure (see also \protect\cite{BucDel95PRL,BucDelZak02PR} for analogous
  states in driven hydrogen atoms). \label{hsN8}}
\end{figure*}

Quite naturally, for the same field amplitude and frequency as above,
the diabatic continuation of the lowest state $n = 1$ of the $N = 8$
series represents the Floquet state that is anchored to the remainder
of the unperturbed regular structure of the frozen-planet
configuration.
Fig.~\ref{hsN8}(a1 -- a3) shows its Husimi distribution which is
fully localized on the intrinsic island and remains nearly stationary
over one field period. 
Furthermore, Floquet states corresponding to higher quantum numbers $n
> 2$ of the unperturbed system are entirely associated with the
chaotic phase space domain.
As an example, Fig.~\ref{hsN8}(b1 -- b3) shows the Husimi density of the
diabatic continuation of the state $n = 3$, which is dominantly
localized on the unstable fixed point of the 1:1 resonance
(compare Figs.~\ref{hsN8}(b3) and (c3)), and along the associated
separatrix layer.

\begin{figure*}
\begin{center}
\leavevmode
\epsfxsize15cm
\epsfbox{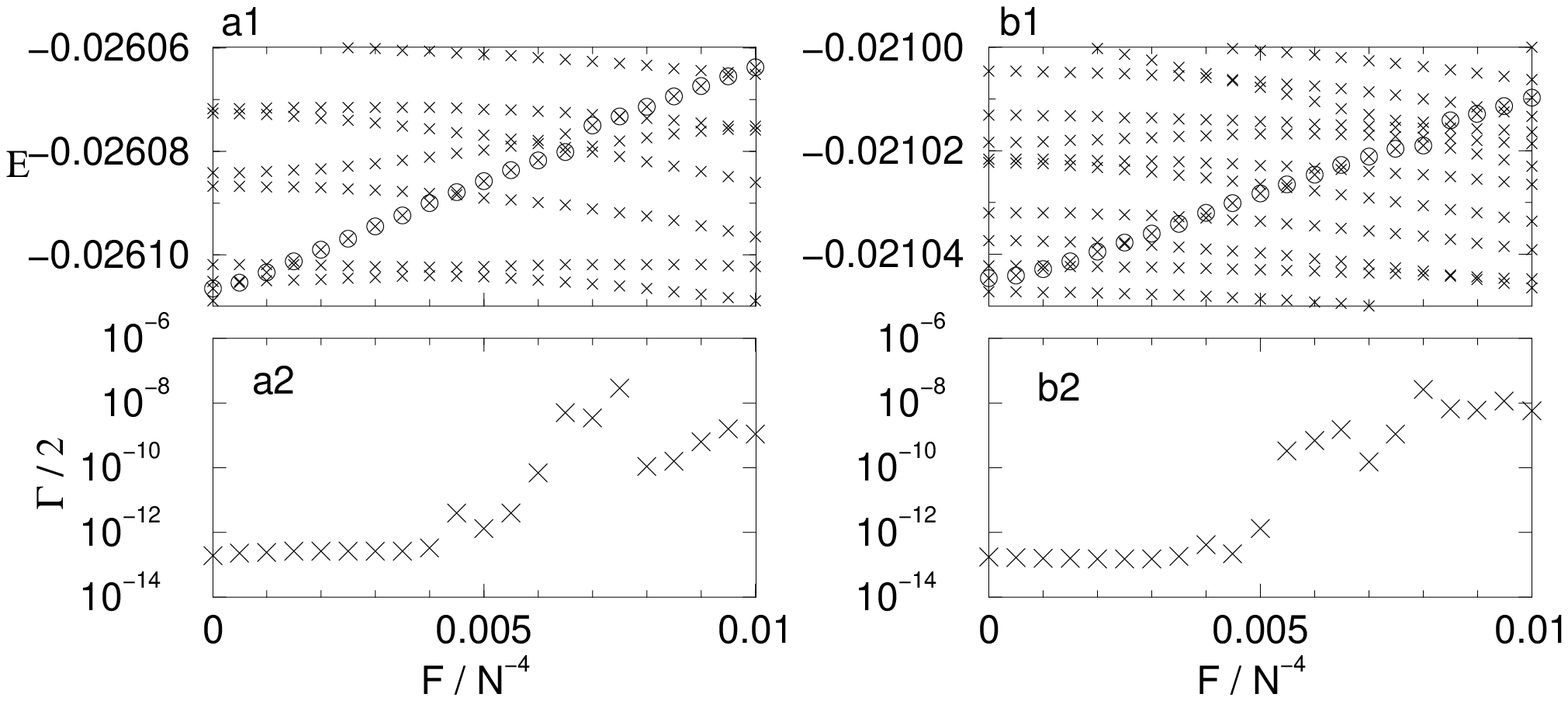}
\end{center}
\caption{Energies and widths of the nondispersive wave packet at 
  $\omega = 0.2 \, N^{-3}$ for $N = 9$ (a1, a2) and       $N = 10$ (b1, b2),
  as a function of the scaled field amplitude $F / N^{-4}$.
  (a1, b1) Evolution of the quasienergies $E$ in the Floquet spectrum.
  The 1:1 resonance state (marked by circles) diabatically originates 
  from the unperturbed state $n = 2$ at $N = 9$, and from the state 
  $n = 3$ at $N = 10$.
  As for $N = 8$ (see Fig.~\ref{flsp}a), its quasienergy undergoes a
  significant shift towards higher values of $E$ with almost
  constant slope, what indicates the pronounced wave packet character
  of this eigenstate in configuration space.
  (a2, b2) Ionization widths of the wave packet states.
  As in Fig.~\ref{flsp}(c), local, sharp enhancements of $\Gamma$
  occur near avoided crossings with higher-lying states of the same
  series, which are associated with the chaotic phase space domain.
  Notice that the precise positions and strengths of the avoided crossings
  and, consequently, their influence on the wave packet's lifetime at a
  given scaled field amplitude $F / N^{-4}$, can be completely different
  for $N$ (a1, a2) and $N + 1$ (b1, b2).
\label{flsp9}}
\end{figure*}

A similar scenario is encountered for different quantum numbers $N$ of
the inner electron (with adjusted field parameters $\omega = 0.2 \,
N^{-3}$ and $F = 0.005 \, N^{-4}$).
Starting from $N = 5$, we find, for each $N$, one Floquet state that is
localized on the intrinsic island (the diabatic continuation of the
island's ground state in the unperturbed spectrum), as well as one
nondispersive wave packet state localized on the 1:1 resonance island,
represented by the diabatic continuation of the second unperturbed
state for $N < 10$, and by the continuation of the third or a higher
state for $N \geq 10$.
As for $N = 8$, these wave packet states are characterized by a
pronounced, almost linear level shift with increasing $F$ towards
higher quasienergies (Fig.~\ref{flsp9}). 

\begin{figure*}
\begin{center}
\leavevmode
\epsfxsize15cm
\epsfbox{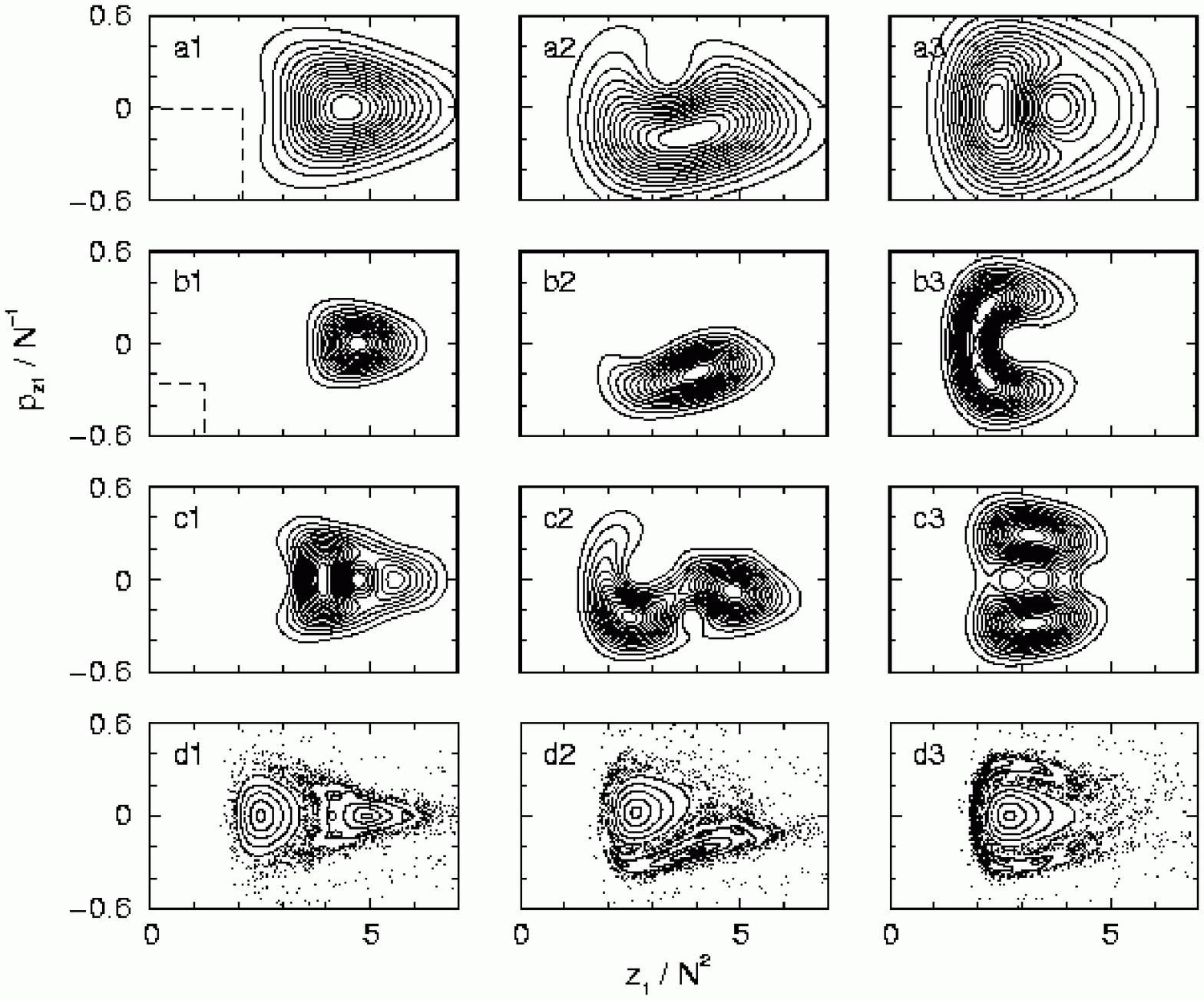}
\end{center}
\caption{Nondispersive two-electron wave packet for $N = 5$ (a1 -- a3)
  and $N = 15$ (b1 -- b3), as well as the first excited mode of the
  nondispersive wave packet for $N = 15$ (c1 -- c3), at $\omega = 0.2
  \, N^{-3}$ and $F = 0.005 \, N^{-4}$.
  The Husimi density of the Floquet states is plotted for the phases
  (a1 -- d1) $\omega t = 0$, (a2 -- d2) $\pi / 2$,
  (a3 -- d3) $\pi$, and compared to the corresponding
  classical Poincar\'{e} sections (d1 -- d3).
  With increasing $N$, the wave packet is more and more localized
  on the center of the 1:1 resonance, due to the decreasing relative
  size of $2 \pi \hbar$ (as compared to the typical atomic scales, see
  Eqs.~(\ref{scr})) indicated by the dashed rectangles in
  (a1) and (b1).
  At $N = 15$, the contour lines of the Husimi density reproduce the
  structure of the island in the classical Poincar\'{e} section rather
  well.
\label{hsN}}
\end{figure*}

Fig.~\ref{hsN} compares the Husimi distributions of the wave packet
states for $N = 5$ and $N = 15$.
Since the phase space area of the island increases linearly with $N$
while $\hbar$ remains constant, the electronic density is, with
increasing $N$, more and more localized around the center of the 1:1
resonance island. 
At $N = 15$, the contour lines of the Husimi density reproduce the
classical island structure in the Poincar\'{e} section already very
well.
In fact, at such high values of $N$, the island is already large
enough to trap an additional wave packet state, giving rise to a
first excited mode of the nondispersive wave packet (see
Fig.~\ref{hsN}(c1 -- c3), and Ref.~\cite{ZakDelBuc95PRL} for an
analogous case in driven hydrogen).

\subsection{Lifetimes of the wave packet states}

\label{LW}

As already mentioned in the Introduction, the time scale over which the
nondispersive wave packet follows the resonantly driven classical
orbit without spreading is only limited by the finite ionization rate
of the corresponding Floquet state.
Since this ionization proceeds via a classically forbidden tunneling
process, the associated decay rate is, as in driven hydrogen
\cite{Buc93,BucDel95PRL,ZakDelBuc95PRL}, very small.
For the wave packet state of the $N = 8$ series ($\omega = 0.2 \, N^{-3}$ and
$F = 0.005 \, N^{-4}$) for instance, we obtain $\Gamma / 2  \simeq
1.77 \cdot 10^{-11}$ a.u., which 
corresponds to a lifetime of $\tau \simeq 1.8 \cdot 10^6$ field cycles
before the wave packet population is appreciably redistributed over
the ionization continuum.

\begin{figure*}
\begin{center}
\leavevmode
\epsfxsize15cm
\epsfbox{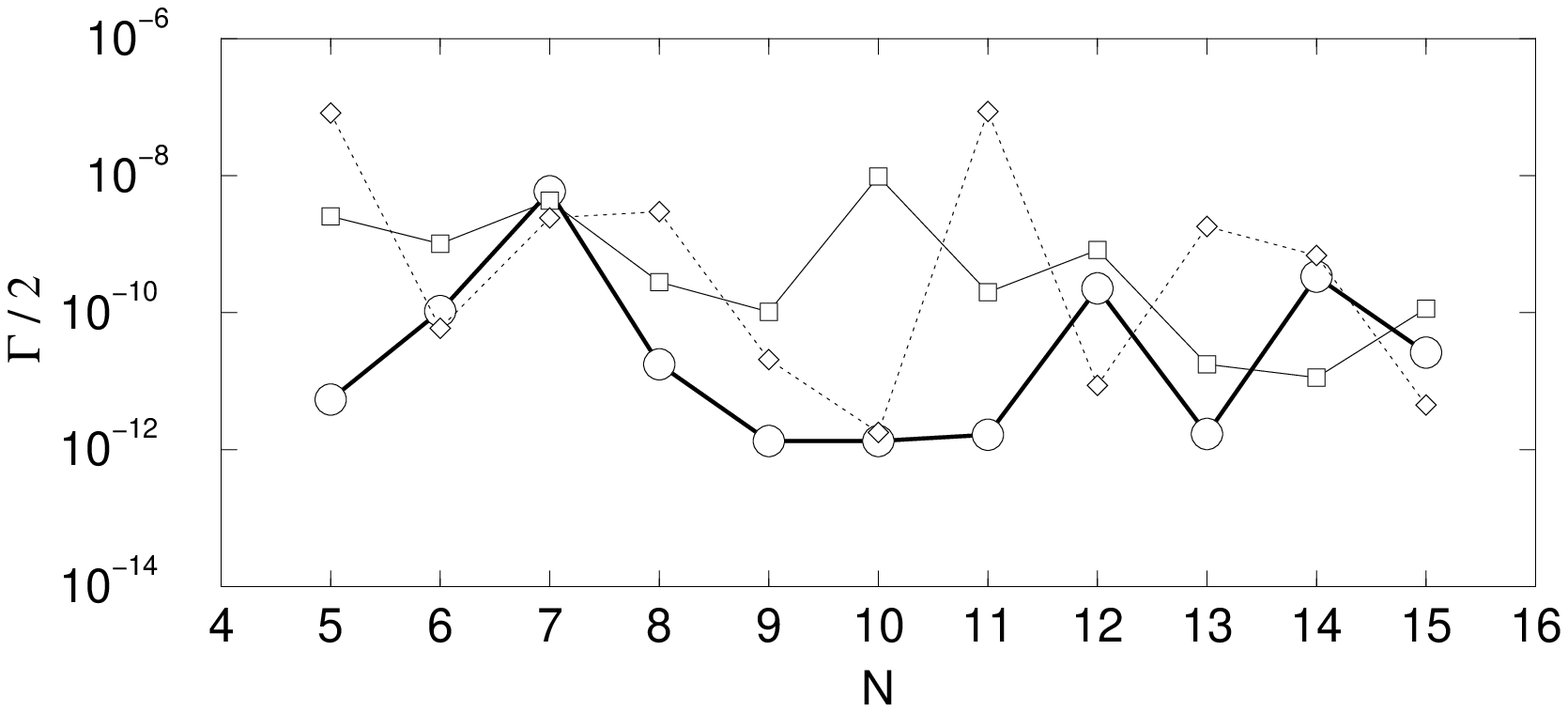}
\end{center}
\caption{Ionization rates (HWHM) of the wave packet state localized on
  the 1:1 resonance island, as a function of the inner electron's
  quantum number $N$ ($\omega = 0.2 \, N^{-3}$ and $F = 0.005 \,
  N^{-4}$), for different strengths $F_{st}$ of the static field.
  Circles connected by thick solid lines: $F_{st} = 0$;
  squares connected by thin solid lines: $F_{st} = 0.001 \, N^{-4}$;
  diamonds connected by dotted lines: $F_{st} = 0.002 \, N^{-4}$.
  The strong fluctuations of the rates are attributed to the coupling
  to Floquet states that are associated with the chaotic domain of
  phase space.
  \label{wpwidths}}
\end{figure*}

In general, one expects the ionization rate of the wave packets to
decrease on average exponentially $\sim \exp( - S N )$ with $N$, where
$S$ is the semiclassical action across the tunneling barrier that
confines the resonance island at $N = 1$.
In contrast to the unperturbed collinear configuration discussed in
Section \ref{1D} (Fig.~\ref{enwd}(c)), however, such 
an exponential decrease is {\em not}
observed in our calculations.
As we see in Fig.~\ref{wpwidths}, the ionization rates exhibit, in the
range $N = 5 \ldots 15$, strong, quasi-erratic fluctuations between
$\Gamma/2 \sim 10^{-12}$ a.u. and $\Gamma/2 \sim 10^{-8}$ a.u.,
corresponding to lifetimes between $10^4$ and $10^7$ field cycles.
No significant tendency towards lower values for increasing $N$ can be
unambiguously identified.

The observed fluctuations of $\Gamma$ represent a clear signature of the mixed 
regular-chaotic structure of the classical phase space (see, e.g.,
Fig.~\ref{hsN}(d1--d3)).
In contrast to globally regular systems such as the unperturbed $Zee$
configuration (see Section \ref{1Dw}), the coupling from the locally
bound domain to the ionization continuum is {\em not} given by a
single tunneling rate.
Instead, it is mediated by a {\em two-step} process which, besides 
tunneling through the barriers of the regular island, involves also 
chaotic transport through irregular phase space outside the island
towards the continuum.
It is in fact this latter, chaotic component of the coupling process
that generates the fluctuations of the decay rates, on top of their
on average exponential decrease with $N$.
This {\em chaos-assisted} tunneling mechanism, originally formulated
for tunneling transitions between symmetric regular islands in bounded
systems \cite{BohTomUll93PR}, was indeed proven to govern
the decay of nondispersive wave packets in driven one-electron atoms
\cite{ZakDelBuc98PRE}, the latter exhibiting similar quasi-erratic
deviations of $\Gamma$ from a purely exponential law (see also
\cite{HorBuc98EPL}).

In the Floquet spectrum, the chaos-induced modification of the 
tunneling process can essentially be understood via the influence of 
near-resonant chaotic states on the ionization rate of the wave packet
state.
As we see in Fig.~\ref{flsp}(b--d), the widths of the energetically
low-lying members of the $N = 8$ series generally increase with
increasing $F$.
Local, pronounced enhancements of $\Gamma$, however, occur near
avoided crossings (corresponding to multiphoton resonances) with
higher-lying states of the same series, which are associated with the
chaotic phase space domain (compare Fig.~\ref{flsp}(b--d) with
Fig.~\ref{flsp}(a)).
Hence, at a given scaled field amplitude, e.g.\ $F = 0.005 N^{-4}$,
the decay rate of the wave packet state depends sensitively on the 
effective interaction with nearby chaotic states in the spectrum,
the latter being essentially determined by the energetic distance to 
those states, as well as by the corresponding coupling matrix elements.
Since these ingredients vary in a random-like way with the quantum
number $N$ of the inner electron (at fixed $F / N^{-4}$), the
resulting effect on the wave packet's ionization rate may turn
completely different when changing $N$ to $N + 1$ (compare here
Fig.~\ref{flsp}(a,b) to Figs.~\ref{flsp9}(a1,a2) and
(b1,b2)), and therefore cannot be predicted by simple
semiclassical considerations.

The average exponential decrease of the ionization rate, on the other
hand, should in principle be amenable to semiclassical estimations, as
it is essentially determined by the height of the tunneling barrier
that confines the resonance island in phase space.
Semiclassical approaches to quantitatively evaluate the associated
tunneling rates for nonintegrable systems are presently under
construction (see \cite{BroSchUll01PRL}).\footnote{Also note,
that the critical field amplitude needed to induce an appreciable increase of
the decay rates beyond their minimum value at small field amplitudes coincides
remarkably well for $N=8,9,10$, as equally observable in Figs.~\ref{flsp} and
\ref{flsp9}. This is a signature of the transition from globally regular to
dominantly chaotic dynamics in the classical phase space, and highlights the
relevance of the classical scaling laws (\ref{scr}), also for the quantum
mechanical ionization process.}

\subsection{Influence of an additional, static electric field}

\label{SF}

\begin{figure*}
\begin{center}
\leavevmode
\epsfxsize15cm
\epsfbox{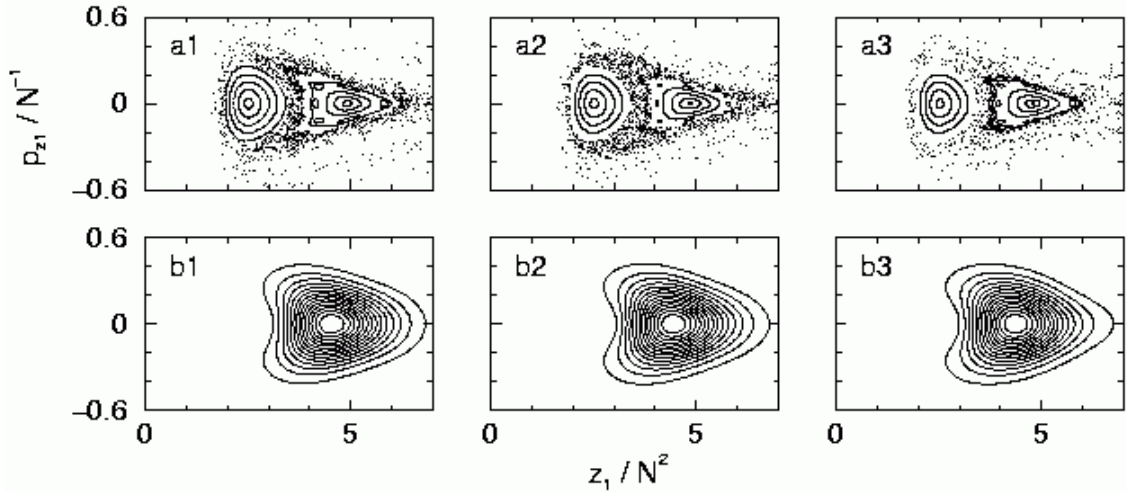}
\end{center}
\caption{(a1 -- a3) Poincar\'{e} section of the classical dynamics at
  $\omega = 0.2$ and $F = 0.005$, with a static field $F_{st} = 0$
  (a1), $0.001$ (a2), $0.002$ (a3). 
  (b1 -- b3) Husimi density of the wave packet state localized on the
  1:1 resonance island for $N = 8$, at $\omega = 0.2 \, N^{-3}$ and $F
  = 0.005 \, N^{-4}$, calculated in presence of a static field
  $F_{st} = 0$ (b1), $0.001 \, N^{-4}$ (b2), 
  $0.002 \, N^{-4}$ (b3) (at driving phase $\omega t = 0$).
  The phase space projection of the wave packet remains almost
  invariant when increasing the static field from $F_{st} = 0$
  to $0.002 \, N^{-4}$, in agreement with the observation
  that also the classical 1:1 resonance island is largely unaffected
  by the additional perturbation.
  \label{ndwpst}}
\end{figure*}

Let us finally discuss to which extent the wave packet dynamics of the
collinear atom is modified by the application of an additional, static
electric field with a strength of up to 20 percent of the driving
field amplitude.
As already pointed out in Section \ref{SD}, such a static field is
required to stabilize the driving-induced nonlinear
resonances in the classical dynamics of the three-dimensional atom.
While the transverse dynamics of the driven configuration is therefore
substantially modified by the static field, the structure of the
collinear phase space remains nearly unaffected.
This is seen in Fig.~\ref{ndwpst}(a1 -- a3) which shows Poincar\'{e} 
sections of the collinear phase space at $\omega = 0.2$ and $F =
0.005$, for gradually increasing static field components $F_{st} =
0 \ldots 0.002$.
Except for a slight reduction of their size, the two dominant regular
islands are not significantly affected by the static perturbation.

Accordingly, we find that the phase space projections of the Floquet
states localized on these islands remain essentially unmodified, too, 
in presence of the static field.
This is illustrated in Fig.~\ref{ndwpst}(b1 -- b3) which shows the 
resulting Husimi density of the nondispersive wave packet state of the
$N = 8$ series at $\omega = 0.2 \, N^{-3}$ and $F = 0.005 \, N^{-4}$.
However, while no visible change occurs in the phase space
localization properties, the ionization rate is significantly affected
by the static field: it increases from 
$\Gamma / 2  \simeq 1.77 \cdot 10^{-11}$ a.u.\ at $F_{st} = 0$, to
$\Gamma / 2  \simeq 3.05 \cdot 10^{-9}$ a.u.\ at $F_{st} = 0.002 \,
N^{-4}$, corresponding to a decrease of the wave packet's lifetime
from $1.8 \cdot 10^6$ to $1.0 \cdot 10^4$ field cycles.
Since the static field notably affects the asymptotic part of the
effective potential experienced by the outer electron (see
Fig.~\ref{adiab}), and thereby modifies the transport towards the
continuum, such an enhancement of $\Gamma$ with $F_{st}$ is not
unexpected.
However, Fig.~\ref{wpwidths} shows that the ionization rates do not
always increase with increasing $F_{st}$.
Instead, a rather erratic dependence of $\Gamma$ on $F_{st}$ and $N$
is encountered, what calls for a more quantitative (and possibly
statistical) analysis of the problem.

\section{Conclusion}

Our present study provides abundant evidence for the stabilizing effect of
classical nonlinear resonances on periodically driven quantum systems, in the
presence of strong electron-electron interactions. We have seen that such
nonlinear resonances, which actually reflect a phase locking phenomenon
\cite{BucDelZak02PR} between the unperturbed quantum dynamics and the driving
field, allow to launch non-dispersive two-electron wave packets along the
highly correlated frozen planet orbit of collinear helium, with tiny,
tunneling induced leakage towards the atomic continuum. As already observed in
simpler one electron systems under external driving, these wave packets can be
identified by the characteristic parameter dependence of their energies and
decay rates, with strong signatures of chaos assisted ionization. Our 1D
quantum calculations corroborate an intimate quantum-classical correspondence,
already at rather weak excitations ($N\geq 5$) of the inner electron, even in
the presence of a static electric field component added to the periodic drive
(which confines the real 3D dynamics to the vicinity of the field polarization
axis, in classical 3D calculations). 
This strongly suggests 
that nondispersive two-electron wave packets exist also in driven 3D helium
(for which, so far, only classical simulations are feasible),
and that similar stabilization phenomena are possible also 
in other,
high dimensional atomic or molecular systems under external forcing. 
However,
our comparison of the life times of 1D and 3D collinear frozen planet states
in the absence of any external perturbation also suggest that the life times
of nondispersive wave packets dramatically depend on the accessible phase
space dimensions, as a direct consequence of the interelectronic repulsion --
absent in single electron Rydberg dynamics. It remains to be seen whether
opening one single transverse degree of freedom -- by extending configuration
space from the $z$-axis to the plane, in a 2D model (now within reach for most
advanced supercomputing facilities) --  suffices to restablish the order of
magnitude of the (field free) 3D decay rates, and whether the confinement by
the static field has a systematic effect on the rates in 2D or 3D (with or
without an external drive).

\section{Acknowledgment}
It is a pleasure to acknowledge fruitful and inspiring discussions with
Dominique Delande, Beno\^{\i}t Gr{\'e}maud, Laurent Hilico, Andreas Krug, and
Javier Madro{\~n}ero, as
well as financial support by the DAAD through the PROCOPE program.

\end{document}